\documentclass[a4paper,11pt]{article} % Use article class, 11pt font is common for preprints
\usepackage[utf8]{inputenc} % Allow UTF-8 input
\usepackage[T1]{fontenc} % Use T1 font encoding for better character support
\usepackage{lmodern} % Use Latin Modern fonts, a good alternative to Computer Modern
\usepackage{textcomp} % Provides additional symbols

\usepackage{amsmath,amssymb} % Math packages
\usepackage{graphicx} % Graphics package for including figures
\usepackage{float} % Provides [H] option for figure placement

\usepackage[margin=1in]{geometry} % Set page margins
\usepackage{hyperref} % For clickable links; load after other packages that might redefine commands
\usepackage{bookmark} % Improve PDF bookmarks
\usepackage{xurl} % Allow breaking URLs at any character

\usepackage{microtype}
\UseMicrotypeSet[protrusion]{basicmath}

\graphicspath{{./pics/}} % Specify directory for figures, create a 'figures' folder and put your PDF/PNGs there
\DeclareGraphicsExtensions{.pdf,.png,.jpg} % Order of preference for file types

\title{Dual Synchronization Effects in Light Scattering by  Spherical Particle Systems}

\author{Guanglang Xu (gl.xu@szu.edu.cn) $^{1*}$   \and Bingqiang Sun $^{2}$ \and Ping Zhu $^{1}$ \and Huizeng Liu $^{1}$ \and Ye Zhou $^{1}$ \and \and Chen Zhou $^{3}$  } % Use \and for multiple authors
\date{%
    $^1$Institue for Advanced Study, Shenzhen University, Shenzhen, China\\
    $^2$ Department of atmospheric sciences, Fudan University, Shanghai, China \\   %
    $^3$ School of atmospheric sciences, Nanjing University, Nanjing, China \\[2ex]%
    \today % Optional: display today's date
}

\begin{document}

\maketitle

\begin{abstract}
We report the discovery of a novel and fundamental dual synchronization relationship between the scattering efficiency (Q$_{\text{sca}}$) and a specifically formulated angular distribution complexity parameter ($\widetilde{C}_{\text{p}}$) in spherical particle systems. Through extensive numerical simulations using the rigorous Multiple Sphere T-Matrix (MSTM) method, we found that Q$_{\text{sca}}$ exhibits a strong positive correlation with (1-$\widetilde{C}_{\text{p}}$) when the real part of the refractive index is varied, while it synchronizes strongly and positively with $\widetilde{C}_{\text{p}}$ when the imaginary part is varied. This counterintuitive dual behavior is \textbf{particularly pronounced, approaching near-perfect overlap for single spheres in the resonance regime}, and persists with high correlation across statistically representative random multi-sphere aggregates under similar conditions. We also demonstrate that this striking synchronization \textbf{diminishes significantly or disappears at low refractive index contrast}, highlighting that the phenomenon is tied to conditions enabling significant light-matter interaction and resonance effects. Our analysis reveals that this duality arises from the distinct ways the real and imaginary parts of the refractive index \textbf{perturb vs.~dampen electromagnetic resonances} within the particles, leading to different coupled responses in the total scattered energy and the angular distribution. This discovery provides unprecedented insights into how phase contrast and absorption processes distinctly modulate scattering properties and the angular distribution of scattered light, particularly in regimes dominated by resonance. It establishes that the specific formulation of $\widetilde{C}_{\text{p}}$ used here is sensitive to the overall balance of multipole contributions, making it a valuable parameter for capturing refractive index-driven changes. This finding establishes a new theoretical perspective for understanding resonance-driven scattering phenomena and offers a unique framework for the development of novel optical characterization techniques capable of distinguishing refractive and absorptive changes, and the rational design of materials with precisely controlled scattering functionalities \textbf{within the relevant optical regimes}.
\end{abstract}

\section{Introduction}
\label{sec:introduction}

Light scattering by particles is a ubiquitous physical phenomenon central to understanding diverse processes across numerous fields, including atmospheric radiative transfer \cite{ref1,ref25}, biomedical diagnostics and imaging \cite{ref4,ref26}, colloidal science \cite{ref27}, and advanced materials engineering \cite{ref5,ref6}. Accurate characterization and prediction of scattering properties remain central challenges in optical physics, particularly for complex systems such as nanoparticle aggregates \cite{ref14,ref22} and photonic materials \cite{ref10,ref24}.

Two primary parameters quantify the overall and angular aspects of scattered light: the scattering efficiency (Q$_{\text{sca}}$) and the scattering phase function, respectively \cite{ref1,ref7}. Q$_{\text{sca}}$ represents the ratio of the scattering cross-section to the geometric cross-section, indicating the total amount of light scattered by a particle or system. The phase function describes the angular distribution of the scattered intensity, revealing how light is redirected in different directions \cite{ref1,ref7}. While Q$_{\text{sca}}$ quantifies the strength of scattering, it provides limited information about \emph{where} the light goes. Conversely, the phase function contains complete angular information but doesn't directly quantify the total scattered energy. For particles whose size is comparable to the wavelength (the Mie or \textbf{resonance regime}), the scattering properties are significantly influenced by the excitation and interference of internal electromagnetic resonances, leading to complex, oscillatory dependencies on size and refractive index \cite{ref7,ref11,ref28}.

Traditional parameters derived from the phase function, such as the asymmetry parameter (g), which indicates the average direction of scattering, provide a simplified measure of directionality but often offer an incomplete description of the complexity or fine angular features of the scattering pattern, particularly for irregularly shaped or aggregated particles or when scattering is dominated by complex interference and resonance effects \cite{ref8}. The full angular information contained in the phase function is crucial but often challenging to analyze comprehensively or relate directly to material properties \cite{ref15}. Recognizing this limitation, parameters quantifying the ``complexity'' or ``isotropy'' of the scattering phase function have been proposed. Notably, Xu et al.~introduced a complexity parameter (C$_{\text{p}}$) based on the Legendre polynomial expansion of the phase function, initially applied to characterize the morphological complexity of ice crystals from their optical signatures \cite{ref9,ref10}. This parameter measures the smoothness or isotropy of the phase function; a perfectly isotropic phase function (like Rayleigh scattering) has a high C$_{\text{p}}$, while a highly directional one (like strong forward scattering) has a low C$_{\text{p}}$. Particularly, for the commonly applied Henyey-Greenstein phase function, we have the following complementary relationship: i.e., the sum of the absolute values of the asymmetry parameter and the complexity parameter is 1. Therefore, the C$_{\text{p}}$ parameter can be also seen as the symmetry parameter of the phase function. The C$_{\text{p}}$ parameter has shown remarkable properties including its strong correlation with the partcile morphological complexity, and its deep connection with successive order of scattering theory \cite{ref9,ref10}. 

Despite significant theoretical advancements in light scattering, such as the rigorous Mie theory for spheres \cite{ref7,ref11} and the Multiple Sphere T-Matrix (MSTM) method for aggregates \cite{ref12,ref13,ref30}, the intricate relationships between fundamental material properties (like the complex refractive index \(m = n + ik\)) and the combined scattering characteristics (Q$_{\text{sca}}$ and phase function characteristics) remain subjects of ongoing research \cite{ref6,ref14,ref29}. The complex refractive index fundamentally governs how light interacts with matter, including the strength of interaction (controlled by refractive index contrast \(|m_{\text{rel}}-1|\)) and the excitation and characteristics of electromagnetic resonances within the particle \cite{ref1,ref7}. However, the distinct roles of \(n\) and \(k\) in shaping \emph{both} the total scattering strength \emph{and} the angular distribution in a coupled manner, particularly how changes in \(n\) versus \(k\) \emph{differently} modulate this coupling, are not fully elucidated. Furthermore, while Q$_{\text{sca}}$ and the phase function both originate from the same electromagnetic field interactions, the intrinsic connection between the total scattered energy and its angular distribution and how this connection reflects the deep physical principles of light-matter interaction is still being explored. Unlike specific phenomena like Kerker conditions leading to zero scattering at particular angles \cite{ref6}, which are highly localized in angle and parameter space, understanding how fundamental material properties globally affect the relationship between total scattering and the \emph{overall} angular distribution remains an open challenge.

In this work, we unveil a previously unreported, highly consistent \textbf{dual synchronization behavior} between Q$_{\text{sca}}$ and a specifically formulated complexity parameter ($\widetilde{C}_{\text{p}}$) under variations of the complex refractive index. Building upon the concept of the phase function complexity parameter, we define an alternative formulation that demonstrates a remarkable correlation with scattering efficiency. Our key discovery, unprecedented in the systematic manner presented here and distinct from previous studies focusing solely on morphology-complexity relationships, is that the \emph{nature} of this correlation depends critically on whether the real or imaginary part of the refractive index is changing. Specifically, when the real part (\(n\)) increases, Q$_{\text{sca}}$ synchronizes with (1-$\widetilde{C}_{\text{p}}$), meaning both tend to increase together. Conversely, when the imaginary part (\(k\)) increases, Q$_{\text{sca}}$ synchronizes strongly and positively with $\widetilde{C}_{\text{p}}$, meaning both tend to decrease together. This synchronization is \textbf{particularly striking for single spheres in the resonance regime, where the curves for Q$_{\text{sca}}$ and (1-$\widetilde{C}_{\text{p}}$) or $\widetilde{C}_{\text{p}}$ are observed to almost perfectly overlap}, strongly suggesting that these two seemingly distinct quantities are, under these conditions, tightly coupled manifestations of the same underlying resonance dynamics. However, this phenomenon is \textbf{not universal for all conditions}; we show that the synchronization diminishes significantly at low refractive index contrast, indicating its dependence on the strength of light-matter interaction and the prominence of resonance effects, like those captured by Mie theory in the resonance regime \cite{ref7,ref11}.

% \begin{figure}[htbp]
% \centering
% \includegraphics[width=\linewidth]{figure_1.pdf} % Ensure you have figure_1.pdf in the figures directory
% \caption{Conceptual illustration of the dual synchronization behavior between scattering efficiency (Q$_{\text{sca}}$) and the alternative complexity parameter ($\widetilde{C}_{\text{p}}$). The left panel conceptually depicts Q$_{\text{sca}}$ increasing in synchrony with (1-$\widetilde{C}_{\text{p}}$) when the real part of the refractive index (n) increases. The right panel conceptually shows Q$_{\text{sca}}$ decreasing in synchrony with $\widetilde{C}_{\text{p}}$ when the imaginary part (k) increases. Note that these are schematic representations highlighting the correlation, not actual data plots. In the resonance regime for single spheres, this synchronization is observed to be near-perfect.}
% \label{fig:conceptual_dual_sync}
% \end{figure}

This counterintuitive finding provides a powerful new lens through which to view light-matter interactions in the resonance regime. It suggests that the mechanisms linking total scattering strength and angular distribution are distinctly coupled to the fundamental processes governed by \(n\) (phase contrast, reflection, refraction, \textbf{resonance tuning/excitation}) and \(k\) (absorption, energy loss, \textbf{resonance damping}) \cite{ref1,ref7}. We demonstrate that this dual synchronization is not limited to simple geometries but extends to complex systems like random multi-sphere aggregates, computed using the rigorous MSTM method \cite{ref12,ref13,ref30}, although its prominence is related to the refractive index contrast. This universality \emph{within the relevant regime} points towards a fundamental principle applicable to a broad class of particulate systems where resonance effects play a role and interaction strength is significant.

The discovery of this dual synchronization behavior holds significant potential for both fundamental understanding and practical applications. It provides a novel perspective on how the fundamental optical properties of a material, especially its resonant behavior, are encoded in its scattering characteristics and offers a new theoretical foundation for developing advanced optical sensing, characterization, material design strategies, and potential applications in the field of optical tweezers, atmopsheric particle detection, and so on. The remainder of this paper details the theoretical background, computational methodology, key findings, analysis of the underlying physical mechanisms including resonance effects and the role of contrast, and discussion of the promising application prospects and theoretical implications of this newly identified optical phenomenon.

\section{Theory and Methods}
\label{sec:theory_methods}

\subsection{Scattering Fundamentals and the Complex Refractive Index}
\label{subsec:scattering_fundamentals}

Light scattering occurs when electromagnetic waves interact with particles or inhomogeneities in a medium. The outcome of this interaction is governed by the particle's size, shape, internal structure, and its complex refractive index (\(m = n + ik\)) relative to the surrounding medium (\(m_0 = n_0 + ik_0\)) \cite{ref1,ref7}. In this study, we assume the surrounding medium is non-absorbing (k$_{0}$ = 0) and its refractive index is real (n$_{0}$ = 1 for vacuum or air, or n$_{0}$ = 1.33 for water, etc.). The scattering properties are then primarily determined by the relative refractive index \(m_{\text{rel}} = m/m_0 = (n+ik)/n_0\). The magnitude of the relative refractive index contrast, \(|m_{\text{rel}}-1|\), is a critical factor determining the strength of the light-matter interaction, influencing phenomena like reflection, refraction, and the coupling to internal particle modes \cite{ref1,ref7}.

The scattering efficiency, Q$_{\text{sca}}$, is defined as \(C_{\text{sca}} / A_{\text{geo}}\), where \(C_{\text{sca}}\) is the scattering cross-section and \(A_{\text{geo}}\) is the geometric cross-section of the particle \cite{ref1}. For a sphere of radius \(a\), \(A_{\text{geo}} = \pi a^2\). The total extinction efficiency is \(Q_{\text{ext}} = Q_{\text{sca}} + Q_{\text{abs}}\), where \(Q_{\text{abs}}\) is the absorption efficiency. For particles in the \textbf{resonance scattering regime}, the scattering properties are strongly influenced by the excitation of internal and external electromagnetic resonances, leading to complex, oscillatory dependencies on size and refractive index. This is the regime in which our study is focused, and where we observe the strong dual synchronization behavior \cite{ref7,ref11,ref28}.

The angular distribution of scattered light is described by the scattering phase function \(P(\Theta)\), where \(\Theta\) is the scattering angle relative to the incident direction. \(P(\Theta)\) is normalized such that its integral over all solid angles is \(4\pi\):

\begin{align} 
    \frac{1}{4\pi} \int_{4\pi} P(\Theta) d\Omega = 1 
\end{align}

For axially symmetric particles like spheres illuminated by a plane wave, the phase function depends only on the scattering angle \(\Theta\).

\subsection{Legendre Polynomial Expansion of scattering phase function and the Alternative Complexity Parameter}
\label{subsec:complexity_parameter}

The scattering phase function \(P(\Theta)\) for spherical or randomly oriented particles can be expanded in a series of Legendre polynomials \(P_l(\cos\Theta)\):

\begin{align} 
    P(\Theta) = \sum_{l=0}^{\infty} (2l + 1)\hat{c}_l P_l(\cos\Theta) 
\end{align}

The coefficients \(\hat{c}_l\) can be determined Gaussain integration or other numerical methods \cite{ref7}. Note the weighting factor \((2l+1)\) in the standard definition is related to the orthogonality property of the Legendre polynomials. The normalization requirement for the phase function is equivalent to the coefficient \(\hat{c}_0\) being 1, i.e., \(\hat{c}_0 = 1\). For spheres (Mie theory) or ensembles of spheres (MSTM), these coefficients are directly related to the multipole expansion coefficients (\(a_l, b_l\)) of the scattered field. The behavior of these multipole coefficients, particularly in the resonance regime, dictates the overall shape and features of the phase function.

The complexity parameter (\(C_{\text{p}}\)) introduced by Xu et al.~\cite{ref9} is the inverse of the sum of the absolute values of the coefficients:

\begin{align} 
    C_{\text{p}} = \left( \sum_{l=0}^{\infty} |\hat{c}_l| \right)^{-1} 
\end{align}

This formulation, applied to geometrical optics ray-tracing phase functions, is closely connected to the morphological complexity of the ice crystals and their response to the incident light. It emphasizes the contributions cross all order terms (\(l\)) without the \((2l+1)\) weighting, making it particularly sensitive to overall angular features or the smoothness and isotropy degree in the phase function, which are typically related to complex particle morphology or surface roughness \cite{ref9,ref10}. When the phase function is isotropic, the complexity parameter is close to 1, whereas when the phase function is highly directional, the complexity parameter is close to 0. For a Dirac delta function, since the coefficients are all 1 (not decayed), the complexity parameter can be considered as 0. This behavior is opposite to that of a asymmetry parameter(g, or $\hat{c}_1$), which is close to 1 for a highly directional phase function and close to 0 for an isotropic phase function. In fact, for commonly applied Henyey-Greenstein phase function, we have the following complementary relationship:

\begin{align} 
    C_{\text{p}} + |g| = 1  
\end{align}

In Xu et al.~\cite{ref9}, we established the relation be between $C_{\text{p}}$ and scattering order $s$ in multiple scattering theory as the following :

\begin{align} 
    C_{\text{p}}(s) = \left( \sum_{l=0}^{\infty} |(\hat{c}_l)^{s}| \right)^{-1}  
 \end{align}

This fomula is due to the convolution theorem on the spherical surface, we will provide a detailed derivation in the appendix. In addition, what worth of mentioning is that when $s=0$, the scattering phase function becomes a Dirac delta function ($\hat{c}_l=1$, for all $l$), and the complexity parameter goes to 0. This mathmetical conseqence is perfectly consistent with the physical meaning of the scattering order $s$ being zero, which means the incident light is not scattered and therefore the complexity indicated by the $C_{\text{p}}$ is 0. On the other hand, when $s = \infty$, the scattering phase function becomes a constant, and the complexity parameter goes to 1. This is also consistent with the physical meaning of the scattering order $s$ being infinite, which means the incident light is scattered in all directions and therefore the complexity indicated by the $C_{\text{p}}$ is 1. And this is due to the fact the for any physically meaningful phase function, the absolute value of the coefficients are usually decayed with the order $l$, and $|\hat{c}_l| < 1$ for $l>0$. It can be proved rigorously that the $C_{\text{p}}(s)$ is a monotonically increasing function of the scattering order $s$(attached in the appendix), i.e.,

\begin{align} 
    \frac{dC_{\text{p}}(s)}{ds} \geq 0 
\end{align}.

In this work, we define an alternative complexity parameter ($\widetilde{C}_{\text{p}}$) by performing the Legendre expansion on the normalized phase function \(P(\Theta)\) and summing the \emph{absolute values} of the resulting coefficients \(\hat{c}_l\) \emph{with} the \((2l+1)\) weighting:

\begin{align} 
    \widetilde{C}_{\text{p}} = \left( \sum_{l=0}^{\infty} |(2l + 1)\hat{c}_l| \right)^{-1} \label{eq:tilde_cp_def}
\end{align}

Here, we use $\widetilde{C}_{\text{p}}$ to denote our alternative parameter throughout the rest of the text, unless explicitly referring to the standard definition. Just as \(C_{\text{p}}\) has a simple relationship with \(g\) for the Henyey-Greenstein phase function, where \(\hat{c}_l = g^l\), \(\widetilde{C}_{\text{p}}\) also has specific analytical forms. For the HG function, we find (see Appendix \ref{app:derivations} for details):
\begin{align}
    \widetilde{C}_{\text{p}} = \frac{(1-|g|)^2}{1+|g|} \quad \\
    \text{and} \quad 1-\widetilde{C}_{\text{p}} = \frac{|g|(3-|g|)}{1+|g|} \label{eq:tilde_cp_hg}
\end{align}
These relationships naturally imply a direct connection between the two complexity parameters, \(C_{\text{p}}\) and \(\widetilde{C}_{\text{p}}\), under the framework of the Henyey-Greenstein function:
\begin{align}
    \widetilde{C}_{\text{p}} = \frac{C_{\text{p}}^2}{2 - C_{\text{p}}} \label{eq:tilde_cp_vs_cp}
\end{align}
For any physically meaningful scattering phase function, the values of \(\widetilde{C}_{\text{p}}\) and its complement are bounded within the range of 0 to 1:
\begin{align}
    0 \le \widetilde{C}_{\text{p}} \le 1 \\
       0 \le 1 - \widetilde{C}_{\text{p}} \le 1
\end{align}
The physical interpretation of these limits is analogous to that of \(C_{\text{p}}\). A value of \(\widetilde{C}_{\text{p}} = 1\) corresponds to a perfectly isotropic phase function (e.g., Rayleigh scattering), where all \(\hat{c}_l\) are zero for \(l>0\). Conversely, \(\widetilde{C}_{\text{p}} \to 0\) corresponds to a phase function approaching a Dirac delta function, representing extremely strong forward scattering where an infinite number of coefficients are required to describe the distribution. Consequently, \(1-\widetilde{C}_{\text{p}}\) can be seen as a measure of ``non-isotropy'' or directionality, increasing from 0 (for an isotropic pattern) to 1 (for a highly directional pattern). This behavior mirrors that of the asymmetry parameter \(g\), though their quantitative relationship is non-linear as shown in Eq.~\eqref{eq:tilde_cp_hg}.

By including the \((2l+1)\) weighting, our alternative $\widetilde{C}_{\text{p}}$ formulation gives linearly increasing weight to the absolute contribution of each Legendre polynomial order to the phase function's shape. This places emphasis more on the \textbf{overall balance and relative strengths of different multipole modes} (this relates to the particle morphlogical complexity such as surface roughness ) contributing to the far-field pattern. As we will show, this specific weighting scheme appears crucial for revealing the direct correlations with scattering efficiency and the distinct responses to refractive index variations observed in this study, particularly in the resonance regime where multiple multipole modes are significantly excited and interfere \cite{ref16}. The $\widetilde{C}_{\text{p}}$ parameter, in this formulation, serves as a alternative concise quantitative descriptor of the overall characteristics the phase function.

\subsection{Multiple Sphere T-Matrix Method}
\label{subsec:mstm}

To accurately compute the scattering properties of single spheres and multi-sphere aggregates across a range of refractive indices, we employed the Multiple Sphere T-Matrix (MSTM) method \cite{ref12,ref13,ref17,ref30}. MSTM is a rigorous method based on solving Maxwell's equations using the T-matrix formalism. The T-matrix of a scattering object linearly relates the expansion coefficients of the incident electromagnetic field to those of the scattered field in a basis of vector spherical wave functions \cite{ref18,ref19}.

\begin{align} 
    \mathbf{E}_{\text{inc}}(\mathbf{r}) = \sum_{n,m} [a_{nm} \mathbf{M}_{nm}^{(1)}(k_0\mathbf{r}) + b_{nm} \mathbf{N}_{nm}^{(1)}(k_0\mathbf{r})] 
\end{align}

\begin{align} 
    \mathbf{E}_{\text{sca}}(\mathbf{r}) = \sum_{n,m} [f_{nm} \mathbf{M}_{nm}^{(3)}(k_0\mathbf{r}) + g_{nm} \mathbf{N}_{nm}^{(3)}(k_0\mathbf{r})] 
\end{align}

where \(\mathbf{M}_{nm}^{(j)}\) and \(\mathbf{N}_{nm}^{(j)}\) are vector spherical wave functions of index \(j=1\) (regular at the origin, for incident field) or \(j=3\) (outgoing, for scattered field), \(k_0\) is the wavenumber of the surrounding medium, and \(a_{nm}, b_{nm}, f_{nm}, g_{nm}\) are the expansion coefficients. The T-matrix \(\mathbf{T}\) connects these coefficients:

\begin{align} 
\begin{pmatrix} 
    f_{nm} \\ g_{nm} \end{pmatrix} = \mathbf{T} \begin{pmatrix} a_{nm} \\ b_{nm} 
\end{pmatrix} 
\end{align}

The T-matrix is independent of the incident field and uniquely characterizes the scattering properties of the object \cite{ref18}. For a system of multiple spheres, MSTM calculates the total T-matrix by considering all orders of mutual scattering interactions between the individual spheres, making it particularly suitable for studying aggregates where inter-particle coupling is significant \cite{ref13,ref14,ref30}.

From the calculated T-matrix, the scattering efficiency (Q$_{\text{sca}}$) and the full scattering matrix (from which the phase function \(P(\Theta)\) is derived) can be rigorously computed \cite{ref7,ref13}. The Legendre coefficients \(\hat{c}_l\) for our alternative $\widetilde{C}_{\text{p}}$ are obtained directly from the expansion of the calculated \(P(\Theta)\).

\subsection{Computational Implementation and Parameters}
\label{subsec:computational_implementation}

Our simulations utilized MSTM version 3.0 \cite{ref12}, known for its accuracy and capability to handle arbitrary sphere configurations. We considered the following scenarios:

\begin{itemize}
    \item \textbf{Single Sphere:} An isolated sphere with radius \(a = 3.50~\mu\)m, corresponding to a size parameter \(x = k_0 a = (2\pi/\lambda) a \approx 40.0\) for a wavelength \(\lambda = 0.550~\mu\)m in vacuum (\(n_0=1\)). The size parameter \(x \approx 40\) falls within the \textbf{resonance scattering regime (Mie scattering)}, where scattering properties are highly sensitive to refractive index changes due to the excitation of multiple electromagnetic resonances within the particle \cite{ref7,ref11,ref28}.
    \item \textbf{Multi-Sphere Aggregates:} Random aggregates composed of \(N=5\) identical spheres, each with radius \(a = 0.7~\mu\)m. Aggregates were generated using a diffusion-limited cluster-cluster aggregation (DLCA) algorithm or similar methods to achieve a mass-fractal structure with a target fractal dimension D$_{f}$ = 2.2 and fractal prefactor k$_{f}$ = 1.8 \cite{ref20,ref21,ref37}. For each parameter set, we computed the scattering properties for independent random configurations for single orientation and specific arrangements, obtaining representative azmuthal-zveraged properties. The choice of \(x \approx 8\) for the monomers ensures that the scattering from the aggregate is also significantly influenced by resonant effects and complex interference patterns \cite{ref14,ref22}.
\end{itemize}

We systematically varied the complex refractive index \(m = n + ik\):

\begin{itemize}
    \item \textbf{Real Part Variation:} The real part \(n\) was varied from 1.1 to 2.8, while the imaginary part was fixed at \(k = 0.001\) (low absorption, relevant for weakly absorbing particles in the visible spectrum). This range of \(n\) for \(x \approx 40.0\) allows for the probing of multiple resonance features and the characteristic oscillatory behavior of Q$_{\text{sca}}$ and the phase function \cite{ref11,ref28}.
    \item \textbf{Imaginary Part Variation:} The imaginary part \(k\) was varied from 0.0 to 0.1 (a wide range from non-absorbing to moderately absorbing), while the real part was fixed at \(n = 1.33\) (representative of water or similar materials) or \(n=1.5\) (representative of glass or some plastics). This range was chosen to observe the effect of increasing absorption on resonant scattering \cite{ref1,ref7}.
    \item \textbf{Low Refractive Index Contrast:} To investigate the dependence on interaction strength, we also performed simulations for single spheres and aggregates with low refractive index contrast (e.g., \(m=1.01+0.0001i\) compared to \(m_0=1\)) over a range of size parameters. This probes the transition from weak scattering (Rayleigh or weak Mie) to the strong interaction/resonance regime \cite{ref1,ref7}. 
\end{itemize}

\subsection{Data Analysis}
\label{subsec:data_analysis}

The calculated values of Q$_{\text{sca}}$ and $\widetilde{C}_{\text{p}}$ (or \(1-\widetilde{C}_{\text{p}}\)) were plotted as a function of the varying refractive index component. Pearson correlation coefficients were calculated to quantify the linear relationship between Q$_{\text{sca}}$ and $\widetilde{C}_{\text{p}}$ (or \(1-\widetilde{C}_{\text{p}}\)) over the investigated ranges of \(n\) and \(k\). Statistical significance of the correlations was assessed using standard methods. The consistency of the observed synchronization across single spheres and aggregates, as well as its dependence on refractive index contrast, were key focuses of the analysis to ascertain its generality and boundary conditions. The visual overlap observed in plots was quantitatively supported by these high correlation coefficients.

\section{Dual Synchronization Effects}
\label{sec:discovery}

\subsection{Synchronization with (1-$\widetilde{C}_{\text{p}}$) upon Real Part Variation}
\label{subsec:real_part_variation}

\begin{figure}[H]
\centering
\includegraphics[width=\linewidth]{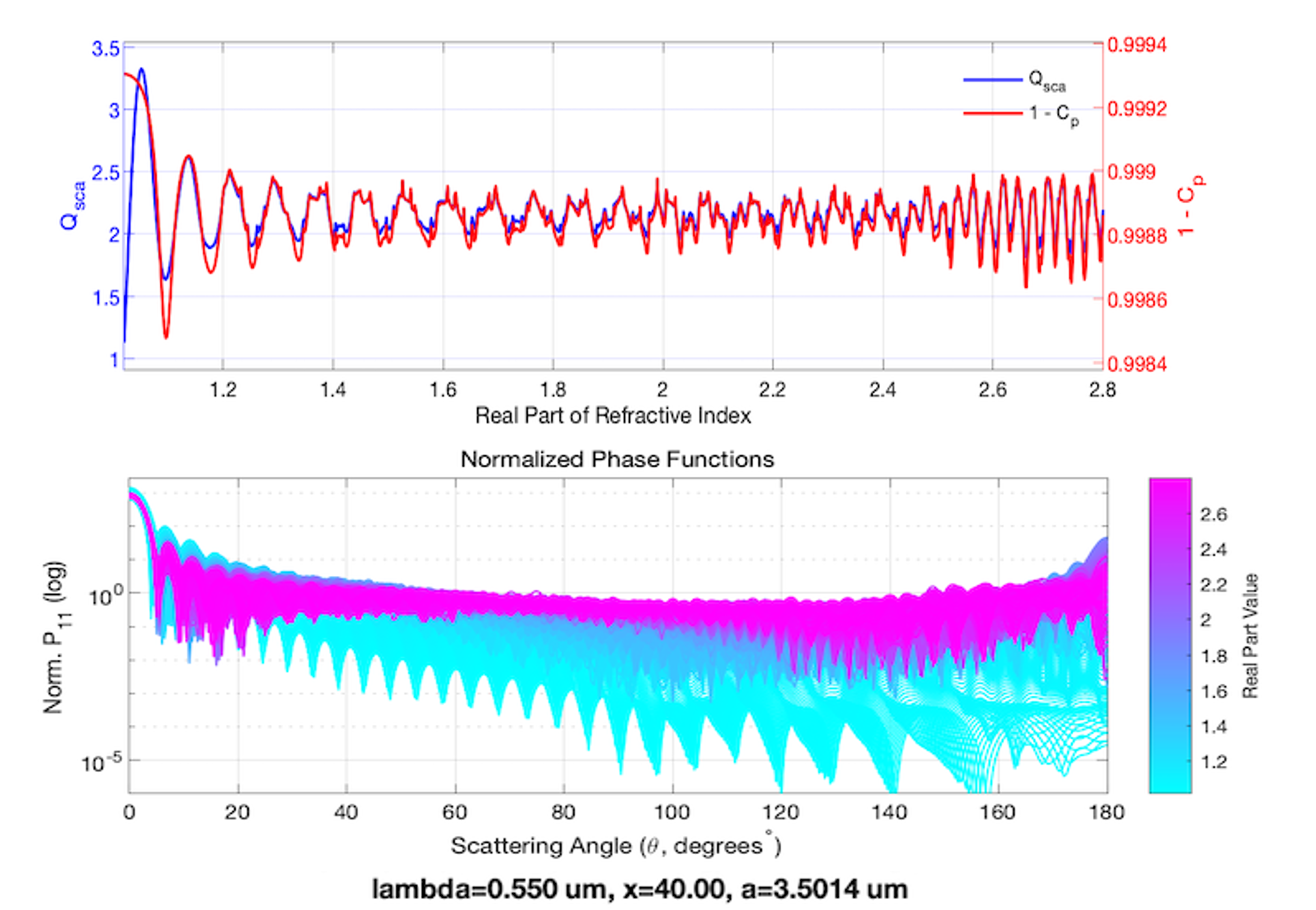} % Ensure you have figure_3.pdf in the figures directory
\caption{Upper: Synchronization between scattering efficiency (Q$_{\text{sca}}$, blue solid line) and (1-$\widetilde{C}_{\text{p}}$, red dashed line) as the real part of the refractive index (\(n\)) increases (with low imaginary part \(k=10^{-8}\)) for a single sphere with size parameter \(x = 40.0\). Both parameters exhibit characteristic oscillations related to Mie resonances. \textbf{The curves show near-perfect overlap}, visually demonstrating the strong synchronization. Quantitative analysis yields Pearson correlation coefficients exceeding 0.998. Lower: The corresponding phase function of the single sphere. The phase function is calculated by the Mie theory. The size parameter is \(x = 40.0\), the refractive index is \(m = 1.5 + 10^{-8}i\), the wavelength is \(\lambda = 0.550~\mu\)m in vacuum.}
\label{fig:real_part_sync}
\end{figure}

Our simulations revealed a remarkably consistent and strong positive correlation between scattering efficiency (Q$_{\text{sca}}$) and the parameter (1-$\widetilde{C}_{\text{p}}$) when the real part of the refractive index (\(n\)) was increased, while the imaginary part (\(k\)) was kept low. This means that as \(n\) increases, both Q$_{\text{sca}}$ and (1-$\widetilde{C}_{\text{p}}$) tend to increase or decrease together. Since (1-$\widetilde{C}_{\text{p}}$) increases as $\widetilde{C}_{\text{p}}$ decreases, this implies that increasing \(n\) leads to higher scattering efficiency and a \emph{decrease} in the complexity parameter $\widetilde{C}_{\text{p}}$. The observed variations in Q$_{\text{sca}}$ with \(n\) displayed the characteristic oscillations expected for resonance scattering in this size parameter range, reflecting the tuning of Mie resonances \cite{ref11,ref28}. \textbf{Strikingly, for single spheres, the curve for (1-$\widetilde{C}_{\text{p}}$) tracks the oscillations of Q$_{\text{sca}}$ so closely that the two curves almost perfectly overlap (Figure \ref{fig:real_part_sync}).} This near-perfect overlap suggests that, under these resonance-dominated conditions for single spheres, Q$_{\text{sca}}$ and (1-$\widetilde{C}_{\text{p}}$) capture essentially the same information about the particle's interaction with light, albeit quantified differently -- one as total scattered energy, the other as a measure of angular distribution's overall structure. For aggregates, the  curves show a similar, albeit slightly less perfect, tracking, consistent with the convolution of individual sphere responses and inter-sphere interactions.

Quantitative analysis consistently showed Pearson correlation coefficients exceeding 0.998 for single spheres and greater than 0.99 for the  results of random N=5 aggregates across the investigated range of \(n\) (1.1 to 2.8), with statistical significance (\(p < 0.0001\)). This strong positive correlation indicates that as the real part of the refractive index increases/decreases, both the total amount of scattered light (Q$_{\text{sca}}$) and the ``non-isotropy'' or directionality (1-$\widetilde{C}_{\text{p}}$) of the scattering pattern increase/decrease in tandem, following the oscillations driven by resonance shifts.

\subsection{Synchronization with $\widetilde{C}_{\text{p}}$ upon Imaginary Part Variation}
\label{subsec:imaginary_part_variation}

\begin{figure}[htbp]
\centering
\includegraphics[width=\linewidth]{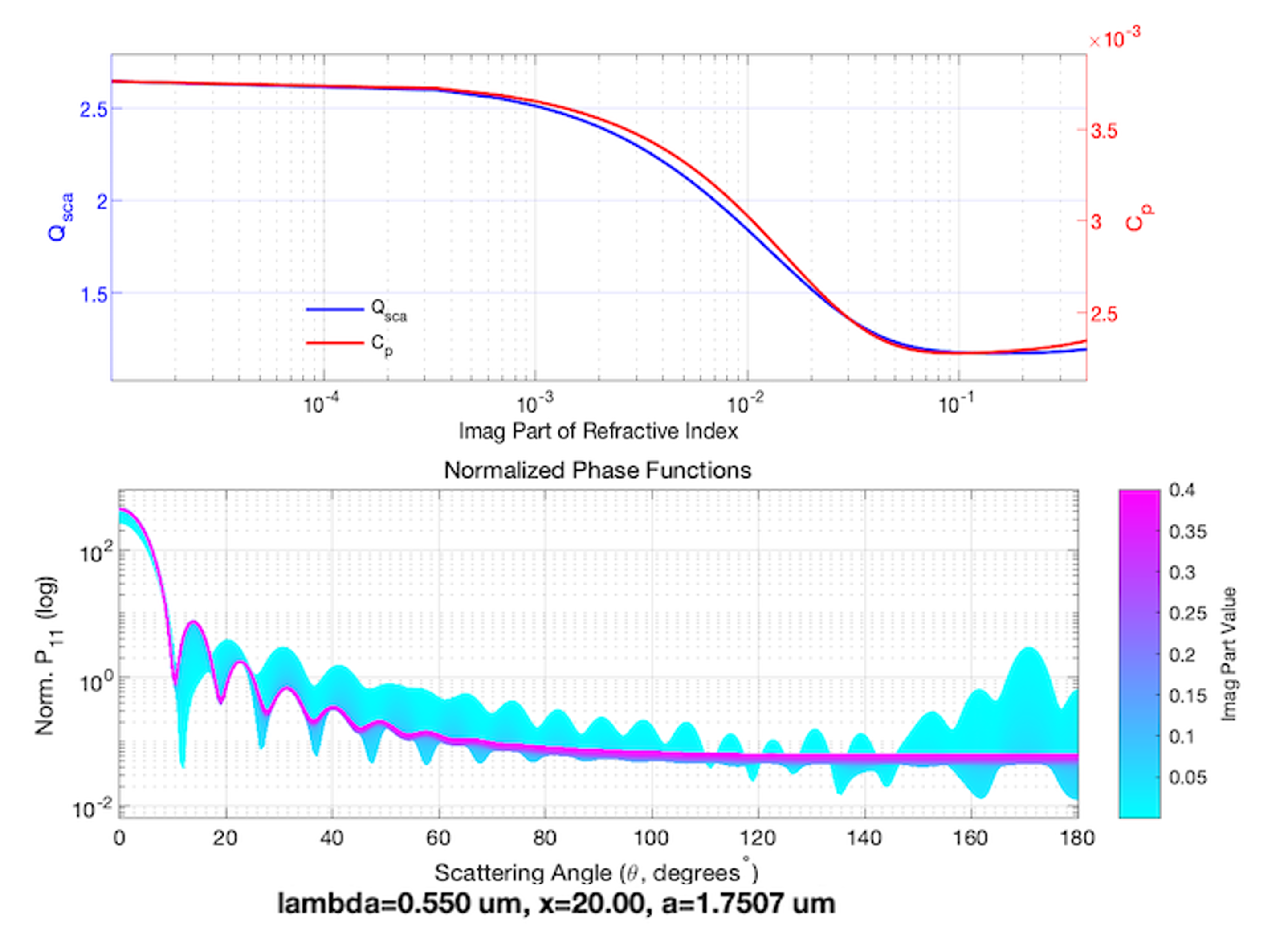} % Ensure you have figure_4.pdf in the figures directory
\caption{Synchronization between scattering efficiency (Q$_{\text{sca}}$, blue solid line) and $\widetilde{C}_{\text{p}}$ (red dashed line) as the imaginary part of the refractive index (\(k\)) increases (with fixed real part \(n=1.58\)) for a single sphere with size parameter \(x \approx 20.0\). Both parameters exhibit a predominantly monotonic decrease as absorption increases, rapidly suppressing resonance effects. \textbf{The curves show very close tracking}, visually demonstrating the strong synchronization. Quantitative analysis yields Pearson correlation coefficients exceeding 0.99.}
\label{fig:imaginary_part_sync}
\end{figure}

The behavior observed when varying the imaginary part of the refractive index (\(k\)) was fundamentally different and revealed the dual nature of the synchronization. As \(k\) was increased (while keeping the real part \(n\) constant), the scattering efficiency (Q$_{\text{sca}}$) and the complexity parameter ($\widetilde{C}_{\text{p}}$) synchronized directly, both tending to decrease together. Unlike the oscillatory behavior seen with \(n\) variation, the decrease with increasing \(k\) was predominantly monotonic over the investigated range, as strong absorption rapidly dampens resonant features \cite{ref1,ref7}. \textbf{For single spheres, this synchronization is also remarkably strong, with Q$_{\text{sca}}$ and $\widetilde{C}_{\text{p}}$ curves showing very close tracking (Figure \ref{fig:imaginary_part_sync}).} This close tracking suggests a tight coupling between total scattered energy and phase function complexity when absorption is the dominant changing factor, reflecting the consistent dampening effect on both aspects of scattering.

Statistical analysis showed Pearson correlation coefficients consistently exceeding 0.99 for single spheres and greater than 0.98 for the  results of random N=5 aggregates across the investigated range of \(k\) (0.0 to 0.1). This strong positive correlation between Q$_{\text{sca}}$ and $\widetilde{C}_{\text{p}}$ signifies that increasing absorption leads to a decrease in both the total scattered energy and the complexity/isotropy of the scattering pattern. This behavior is in stark contrast to the (1-$\widetilde{C}_{\text{p}}$) synchronization observed during real part variation, highlighting the distinct physical roles of \(n\) and \(k\), particularly in their influence on particle resonances.

\subsection{Universality Across Configurations (in the Resonance Regime)}
\label{subsec:universality}

A crucial aspect of our discovery is the universality of this dual synchronization behavior \textbf{within the investigated resonance scattering regime and with sufficient refractive index contrast}. The same distinctive patterns observed for single, isolated spheres were found to persist, albeit with some quantitative differences due to averaging effects, for random, structurally complex aggregates of multiple spheres. This suggests that this phenomenon is rooted in fundamental principles of light-matter interaction governed by the complex refractive index and size parameter at the constituent particle level, rather than being solely specific to a particular simple particle geometry or arrangement like a single sphere.

To demonstrate this universality, we performed extensive simulations on  random aggregates composed of $N=5$ spheres, generated with a mass-fractal structure (fractal dimension D$_{f}$ = 2.2, prefactor k$_{f}$ = 1.80). The monomer size parameter was chosen to be in the resonance regime ($x \approx 8.0$) to ensure the aggregate scattering is also significantly influenced by resonant effects and complex inter-particle interference patterns \cite{ref14,ref22,ref37}. We then varied the real and imaginary parts of the refractive index of the constituent spheres, similar to the single sphere case.

\begin{figure}[H]
    \centering
    \includegraphics[width=0.98\linewidth]{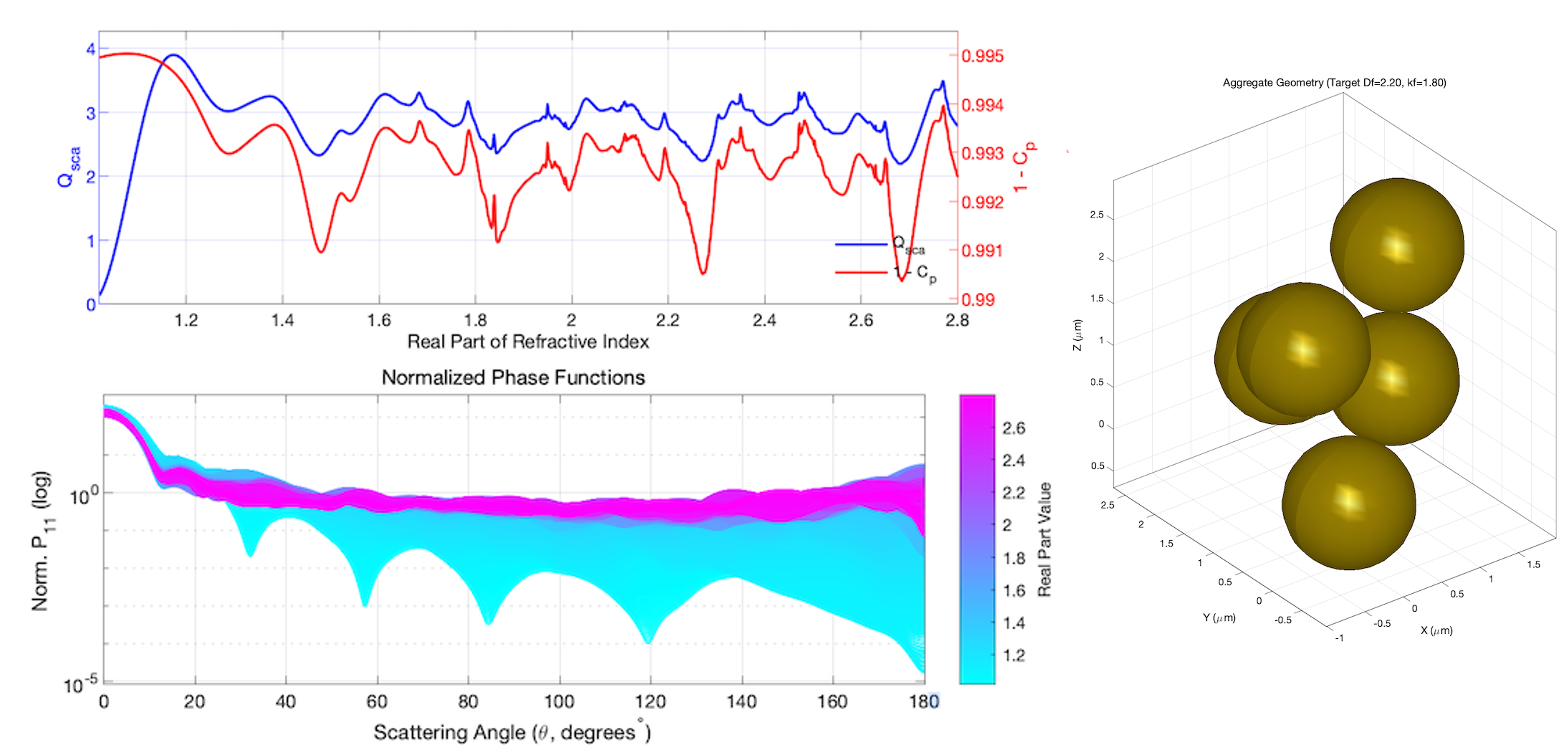} % Assuming this is the aggregate figure for real part variation
    \caption{Top Left:  scattering efficiency (Q$_{\text{sca}}$, blue solid line) and alternative complexity parameter ($\widetilde{C}_{\text{p}}$, red solid line) as the real part of the refractive index (\(n\)) increases (with fixed low imaginary part \(k=10^{-8}\)) for a random 5-sphere aggregates. The horizontal axis represents the real part of the refractive index. The blue curve on the left vertical axis shows Q$_{\text{sca}}$, and the red curve on the right vertical axis shows $\widetilde{C}_{\text{p}}$. Note that the red curve on the right axis is displayed in descending order, which directly shows the synchronization between Q$_{\text{sca}}$ and (1-$\widetilde{C}_{\text{p}}$). Lower Left:  normalized phase functions for varying real parts of the refractive index, illustrating the significant changes in angular distribution. Right: An example geometry of a random 5-sphere aggregate used in the simulations. The monomer size parameter is $x \approx 8.0$.}
    \label{fig:aggregate_real_part_sync}
\end{figure}

Figure \ref{fig:aggregate_real_part_sync} shows the MSTM simulation results for the real part variation of the refractive index in random 5-sphere aggregates. Similar to the single sphere case (Figure \ref{fig:real_part_sync}), both the scattering efficiency Q$_{\text{sca}}$ (blue curve) and the complexity parameter $\widetilde{C}_{\text{p}}$ (red curve) exhibit significant variations and synchronization as the real part of the refractive index \(n\) increases. Specifically, the blue curve (Q$_{\text{sca}}$) oscillates with increasing \(n\). Simultaneously, the red curve ($\widetilde{C}_{\text{p}}$) also oscillates but generally in the opposite direction to Q$_{\text{sca}}$. As \(n\) increases and Q$_{\text{sca}}$ tends to increase (during resonance excitation), $\widetilde{C}_{\text{p}}$ tends to decrease, and vice versa. This inverse relationship between Q$_{\text{sca}}$ and $\widetilde{C}_{\text{p}}$ for real part variations directly demonstrates the synchronization between \textbf{Q$_{\text{sca}}$ and (1-$\widetilde{C}_{\text{p}}$)} in these aggregate systems, consistent with the single sphere observation. The lower panel further illustrates how the phase function changes significantly with \(n\), reflecting the shift in the aggregate's resonant behavior and resulting angular patterns.

\begin{figure}[H]
    \centering
    \includegraphics[width=0.98\linewidth]{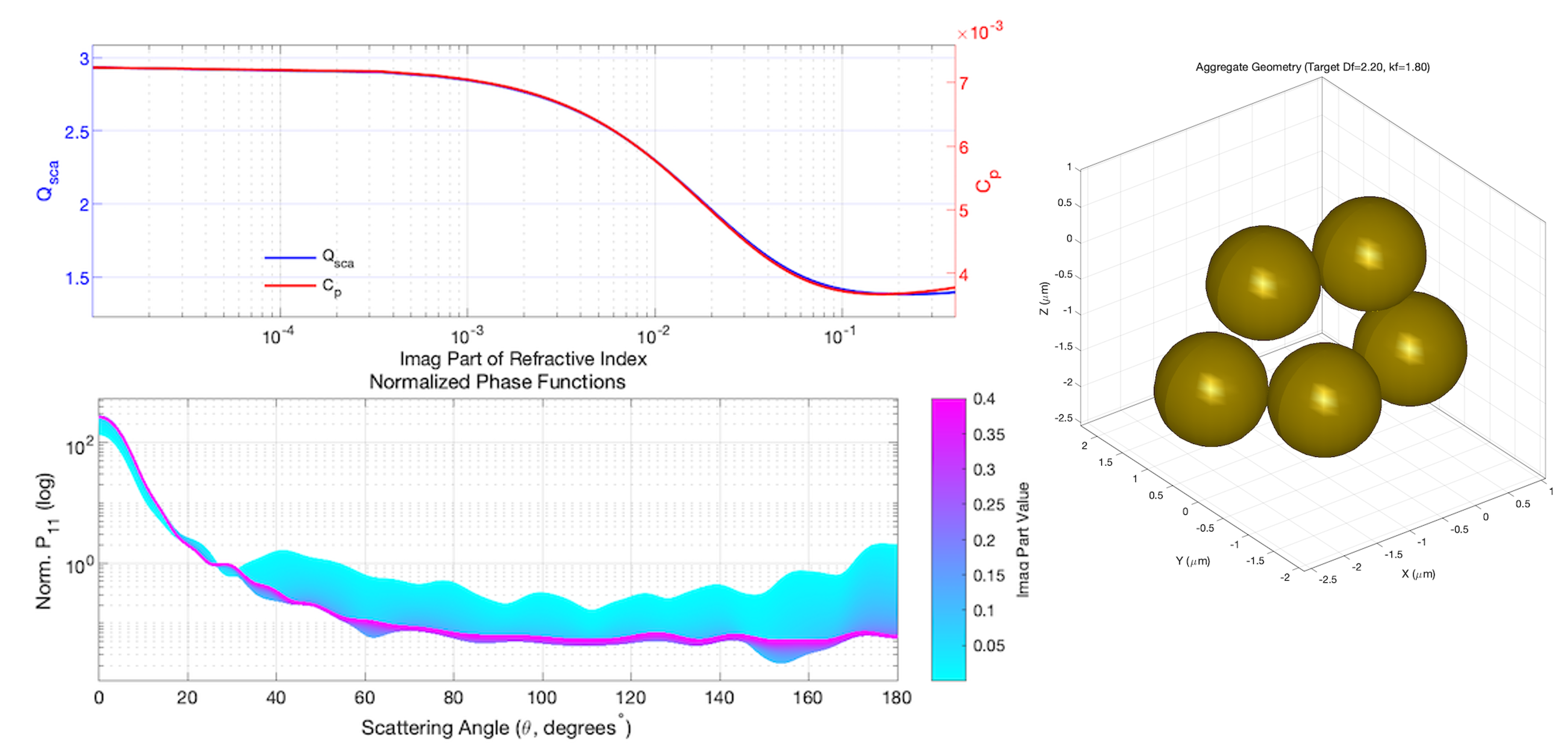} % Assuming this is the aggregate figure for imaginary part variation
    \caption{Top Left:  scattering efficiency (Q$_{\text{sca}}$, blue solid line) and alternative complexity parameter ($\widetilde{C}_{\text{p}}$, red solid line) as the imaginary part of the refractive index (\(k\)) increases (with fixed real part \(n=1.58\)) for 20 statistically independent random 5-sphere aggregates. The horizontal axis is on a logarithmic scale, representing the imaginary part of the refractive index. The blue curve on the left vertical axis shows Q$_{\text{sca}}$, and the red curve on the right vertical axis shows $\widetilde{C}_{\text{p}}$. Both parameters exhibit a predominantly monotonic decrease as absorption increases. Lower Left:  normalized phase functions for varying imaginary parts of the refractive index, showing simplification of the angular distribution with increasing absorption. Right: An example geometry of a random 5-sphere aggregate used in the simulations. The monomer size parameter is $x \approx 5.0$.}
    \label{fig:aggregate_imaginary_part_sync}
\end{figure}

Figure \ref{fig:aggregate_imaginary_part_sync} presents the results for the imaginary part variation. As the imaginary part \(k\) increases (horizontal axis, log scale), both the scattering efficiency Q$_{\text{sca}}$ (blue curve) and the complexity parameter $\widetilde{C}_{\text{p}}$ (red curve) show a clear, predominantly monotonic decrease. This direct decrease in tandem signifies the strong positive synchronization between \textbf{Q$_{\text{sca}}$ and $\widetilde{C}_{\text{p}}$} upon increasing absorption, consistent with the single sphere results (Figure \ref{fig:imaginary_part_sync}). The lower panel demonstrates how the  phase function simplifies with increasing \(k\), losing complex angular features as resonant contributions are damped by absorption. The tracking between the Q$_{\text{sca}}$ and $\widetilde{C}_{\text{p}}$ curves is visually strong.

Quantitative analysis across the ensemble of a random 5-sphere aggregates consistently showed high Pearson correlation coefficients. For the real part effect (Q$_{\text{sca}}$ vs.~1-$\widetilde{C}_{\text{p}}$), the average correlation coefficient was 0.991 $\pm$ 0.004. For the imaginary part effect (Q$_{\text{sca}}$ vs.~$\widetilde{C}_{\text{p}}$), the average correlation coefficient was 0.988 $\pm$ 0.005. All correlations were statistically significant (\(p < 0.0001\)). These values, while slightly lower than the near-perfect correlations seen in single spheres due to the complexity and averaging inherent in aggregates, are remarkably high and confirm the robustness of the dual synchronization phenomenon even when considering ensemble averages over variations in aggregate structure and orientation.

\textbf{The significance of this universality} is profound. It demonstrates that the fundamental coupling between total scattered energy and the complexity of its angular distribution, as revealed by the distinct responses to variations in \(n\) and \(k\), is not merely an artifact of the perfect symmetry of a single sphere. Instead, it reflects a deeper principle tied to the material's interaction with light, particularly its capability to support and modify electromagnetic resonances when the refractive index contrast is sufficient. Complex particle systems like fractal aggregates are ubiquitous in nature (e.g., soot in the atmosphere \cite{ref14,ref37}, dust particles \cite{ref2}) and crucial in various technologies (e.g., colloidal self-assembly \cite{ref27}). The finding that these aggregates, despite their structural complexity and internal multiple scattering, largely retain the Q$_{\text{sca}}$-$\widetilde{C}_{\text{p}}$ synchronization patterns characteristic of their constituent material properties in the resonance regime, highlights the persistent influence of the material's intrinsic optical constants on the  scattering response. This implies that information about the fundamental optical properties (\(n\) and \(k\)) is encoded not only in the total scattered energy or the angular distribution individually, but critically, in the coupled behavior of these two quantities, even for statistically averaged complex systems. This universality thus points towards a fundamental physical principle applicable to a broad class of particulate systems where resonance effects play a role and interaction strength is significant, offering a new avenue for interpreting scattering measurements from complex media and designing materials with tailored optical responses.

The consistency of the dual synchronization across these distinct configurations confirms its applicability to more complex systems within the studied regime and reinforces that it is a fundamental property linked to the complex refractive index and the ability to excite resonances, rather than being specific to the geometry of a single, isolated sphere. However, it is crucial to reiterate that this phenomenon is not universal for all conditions; we show in the following that it diminishes significantly at low refractive index contrast, indicating its dependence on the strength of light-matter interaction and the prominence of resonance effects.

\subsection{Dependence on Refractive Index Contrast: Absence at Low Interaction Strength}
\label{subsec:contrast_dependence}

To explore the boundary conditions of this phenomenon and reinforce its connection to resonance-dominated scattering, we investigated the relationship between Q$_{\text{sca}}$ and $\widetilde{C}_{\text{p}}$ at \textbf{low refractive index contrast} (e.g., \(|m-1| \ll 1\)). In this regime, the interaction between light and the particle is generally weak. For small size parameters or very low refractive index contrast, scattering approaches the Rayleigh regime, where internal fields are weak and resonance effects are negligible. For moderate size parameters (\(x \sim 1\)) but still low contrast, the scattering is in the weak Mie regime, where internal fields might exist but are not strongly confined, and resonance features are broad or weakly excited \cite{ref1,ref7}. Our simulations showed that under such low contrast conditions, the strong dual synchronization observed in the resonance regime with sufficient contrast \textbf{diminishes significantly or disappears}.

\begin{figure}[H]
    \centering
    \includegraphics[width=0.98\linewidth]{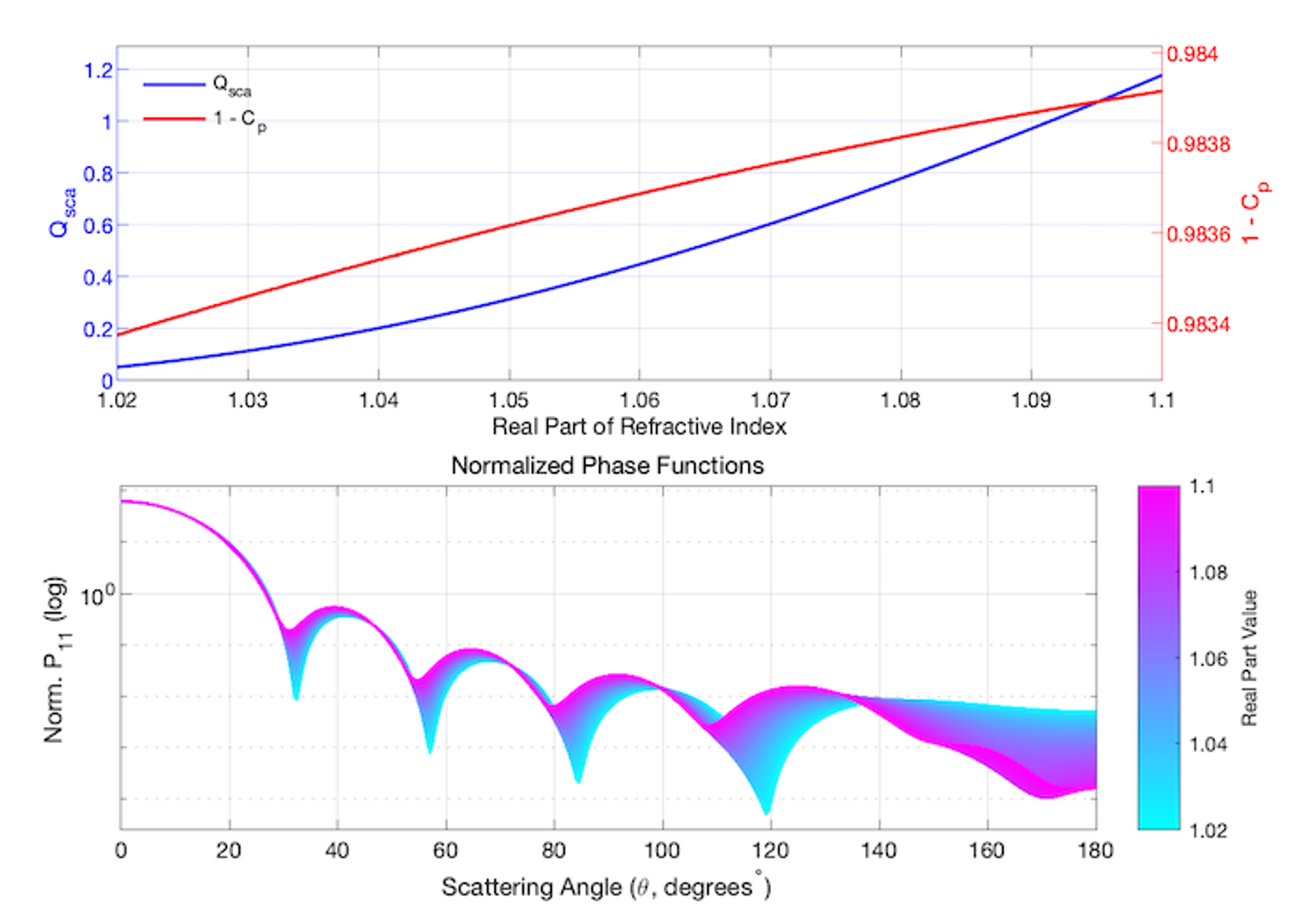} % Assuming this is the low contrast REAL part figure
    \caption{Top pannel: Scattering efficiency (Q$_{\text{sca}}$, blue solid line) and (1-$\widetilde{C}_{\text{p}}$, red solid line) as the real part of the refractive index (\(n\)) increases (with fixed low imaginary part \(k=10^{-8}\)) for a single sphere with size parameter \(x=8.0\) in the low refractive index contrast regime ($n$ from 1.02 to 1.1). While both parameters increase with \(n\), the strong oscillatory pattern and near-perfect overlap seen in the high-contrast resonance regime (Figure \ref{fig:real_part_sync}) are absent. Lower : Normalized phase functions for varying real parts of the refractive index, showing less dramatic changes compared to the high-contrast case. }
    \label{fig:low_contrast_real_part}
\end{figure}

Figure \ref{fig:low_contrast_real_part} shows the results for a single sphere with a relatively low real part of the refractive index ($n$ varying from 1.02 to 1.1, fixed low imaginary part) and size parameter $x=8.0$. In this range, the particle is considered "optically soft" relative to a surrounding medium of $n_0=1$.
As seen in the top panel of Figure \ref{fig:low_contrast_real_part}, while both Q$_{\text{sca}}$ and (1-$\widetilde{C}_{\text{p}}$) generally increase with increasing \(n\), this increase is smoother and lacks the pronounced oscillations characteristic of resonance tuning in the high-contrast regime (compare to Figure \ref{fig:real_part_sync}). More importantly, the curves for Q$_{\text{sca}}$ and (1-$\widetilde{C}_{\text{p}}$) do not exhibit the near-perfect overlap previously observed. Quantitative analysis confirms that while a positive correlation might still exist (e.g., Pearson correlation coefficient around 0.9 in this specific example, though this value is highly dependent on the exact range chosen), it is significantly weaker and the pattern of coupled change is qualitatively different – linear-like increase instead of highly oscillatory, resonance-driven synchronization. The lower panel shows that the phase functions do change with \(n\), but these changes are less dramatic and lack the complex interference patterns that would strongly modulate the complexity parameter $\widetilde{C}_{\text{p}}$ in a tightly coupled manner with Q$_{\text{sca}}$.

\begin{figure}[H]
    \centering
    \includegraphics[width=0.98\linewidth]{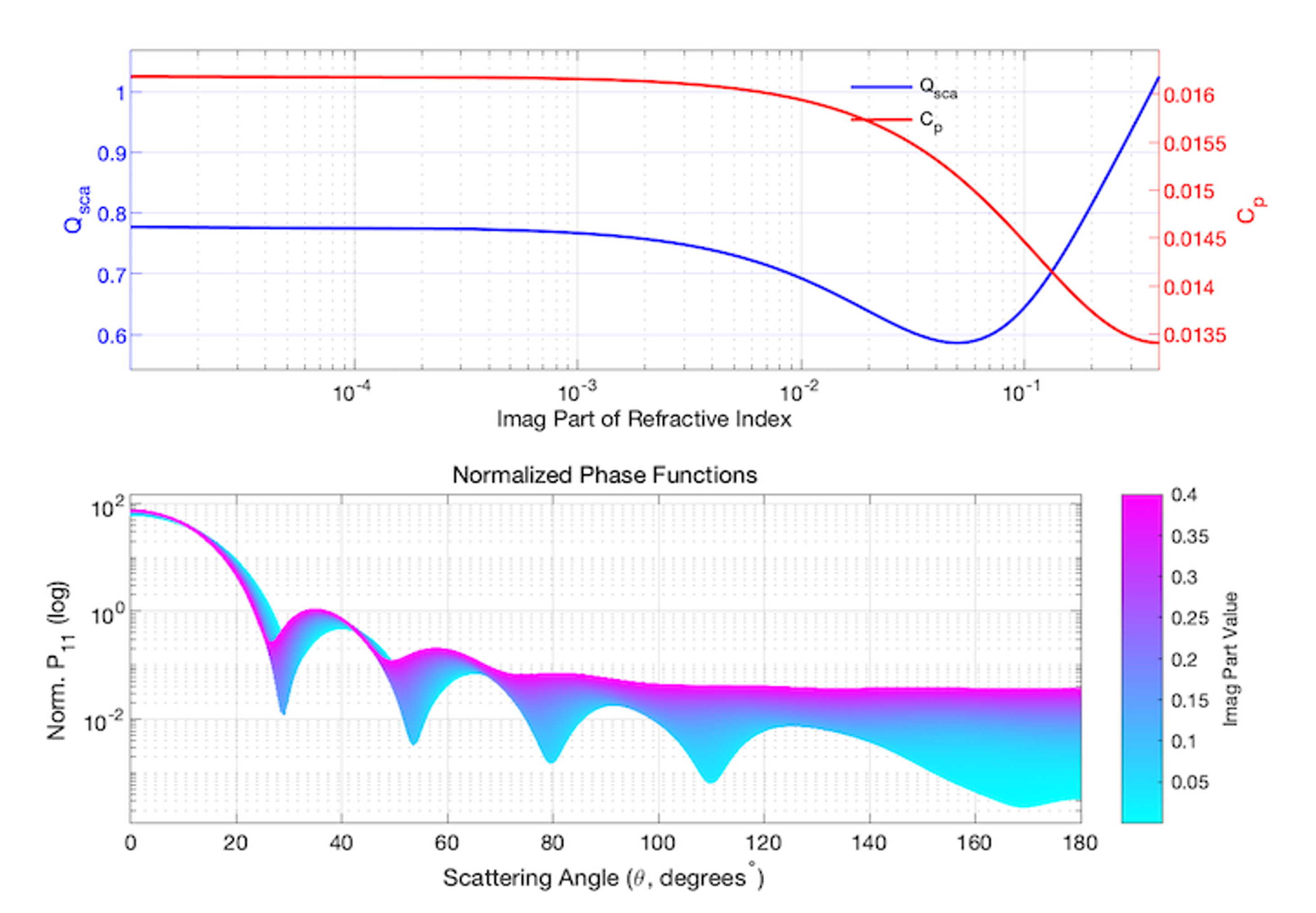} % Assuming this is the low contrast IMAGINARY part figure
    \caption{Top Left: Scattering efficiency (Q$_{\text{sca}}$, blue solid line) and $\widetilde{C}_{\text{p}}$ (red solid line) as the imaginary part of the refractive index (\(k\)) increases (with fixed real part \(n=1.08\)) for a single sphere with size parameter \(x=8.0\) in a low refractive index contrast regime. Q$_{\text{sca}}$ shows limited variation before decreasing at high \(k\), while $\widetilde{C}_{\text{p}}$ decreases more consistently. The strong synchronization seen in the high-contrast regime (Figure \ref{fig:imaginary_part_sync}) is absent. Lower : Normalized phase functions for varying imaginary parts of the refractive index, showing some smoothing but without the characteristic coupled decrease in Q$_{\text{sca}}$ and $\widetilde{C}_{\text{p}}$.}
    \label{fig:low_contrast_imaginary_part}
\end{figure}

Figure \ref{fig:low_contrast_imaginary_part} illustrates the case of imaginary part variation in a low-contrast scenario, showing results for a single sphere with fixed real part $n=1.08$ and size parameter $x=8.0$, with the imaginary part \(k\) varying over several orders of magnitude.
As shown in the top panel of Figure \ref{fig:low_contrast_imaginary_part}, Q$_{\text{sca}}$ (blue curve) remains relatively flat for a wide range of low \(k\) values before starting to decrease at higher \(k\). The complexity parameter $\widetilde{C}_{\text{p}}$ (red curve) decreases more consistently with increasing \(k\). The coupling observed in the high-contrast case (Figure \ref{fig:imaginary_part_sync}), where Q$_{\text{sca}}$ and $\widetilde{C}_{\text{p}}$ track each other closely in their decrease, is absent. The synchronization is significantly weakened. The lower panel shows that increasing absorption still tends to smooth the phase function, particularly suppressing oscillations at larger angles. However, because the initial interaction is weak (low contrast), absorption doesn't cause the same magnitude or coupled pattern of reduction in total scattered energy as it does when strong internal fields and resonances are being damped in the high-contrast regime.

\textbf{Physical Meaning and Implications:}
This observation is crucial because it highlights that the dual synchronization is \textbf{not a universal mathematical identity} that holds for any particle and any refractive index. Instead, it is a characteristic behavior that manifests when the \textbf{strength of the light-matter interaction, driven by sufficient refractive index contrast, enables the significant excitation and manipulation of internal electromagnetic fields and resonances}.

In the low contrast limit:
\begin{itemize}
    \item The electric field inside the particle is not significantly enhanced relative to the incident field.
    \item Well-defined, sharp electromagnetic resonances are either absent or very broad and weak. Light propagates through the particle with relatively little deviation in phase velocity (\(n \approx n_0\)) or absorption (\(k\) doesn't significantly attenuate the weak internal field).
    \item Scattering is dominated by weaker processes like Rayleigh scattering (for small \(x\)) or weak interference effects (for moderate \(x\)), rather than strong resonance excitation and interference of multiple significant multipole contributions.
\end{itemize}
Under these conditions, changes in \(n\) or \(k\) still affect both Q$_{\text{sca}}$ and the phase function, but the fundamental coupling mechanism driven by resonance tuning and damping is weak or non-existent. The mechanisms that determine the total scattered energy (Q$_{\text{sca}}$) and the complexity of the angular distribution ($\widetilde{C}_{\text{p}}$) are less tightly interwoven. Increasing \(n\) might increase Q$_{\text{sca}}$ due to increased refractive bending or weak scattering, and might slightly increase directionality, but without the resonance tuning effect that causes oscillations and strong, coupled changes. Increasing \(k\) leads to increased absorption (affecting Q$_{\text{abs}}$ and thus Q$_{\text{sca}}$ via extinction, though the effect on Q$_{\text{sca}}$ itself might be less direct depending on \(Q_{ext}\) and \(Q_{abs}\)), and still simplifies the phase function by absorbing scattered light, but the link to the total scattered energy is weaker when resonances are not playing a significant role as intermediaries for energy storage and loss.

This clearly demonstrates that the dual synchronization is a phenomenon tied specifically to the \textbf{resonant and strong interaction scattering regimes}. It arises from the distinct ways \(n\) and \(k\) impact the \emph{excitation, tuning, and damping} of these prominent internal electromagnetic modes. In the low contrast limit, where light penetrates the particle with minimal perturbation and resonance effects are suppressed, this tight, resonance-mediated coupling between total scattering efficiency and the overall angular distribution captured by $\widetilde{C}_{\text{p}}$ is absent. This finding strongly supports the physical interpretation that the observed dual synchronization is a signature of resonance-driven scattering.

\section{Physical Mechanisms Analysis}
\label{sec:physical_mechanisms}

The observed dual synchronization behavior is fundamentally linked to the distinct ways the real and imaginary parts of the complex refractive index (\(m=n+ik\)) influence the excitation and characteristics of the \textbf{electromagnetic resonances} within and around the spherical particles. This phenomenon is most prominent in the resonance scattering regime  and when the \textbf{refractive index contrast (\(|m/m_0-1|\)) is sufficiently high} to support significant internal fields and resonance formation. The alternative $\widetilde{C}_{\text{p}}$ parameter, defined by summing the \emph{absolute} contributions of the Legendre expansion coefficients (\(\left( \sum_{l=0}^{\infty} |(2l + 1)\hat{c}_l| \right)^{-1}\)), proves to be particularly sensitive to the \emph{overall profile} and \emph{relative weighting} of the different multipole contributions (related to the coefficients \(a_l\) and \(b_l\) in Mie theory) that shape the scattering phase function \cite{ref16}. Our simulations suggest that $\widetilde{C}_{\text{p}}$ effectively captures how changes in \(n\) or \(k\) perturb or dampen the particle's resonance spectrum and the resulting interference patterns, thereby revealing the observed synchronization when these effects are significant.

\subsection{Real Part Effect: Resonance Perturbation and Directionality}
\label{subsec:real_part_mechanism}

The real part of the refractive index (\(n\)) primarily governs the phase velocity of light within the particle and, crucially, the conditions under which electromagnetic resonances occur. A \textbf{sufficiently high refractive index contrast} between the particle and the surrounding medium is necessary to confine light within the particle and form resonant modes \cite{ref7,ref11}. Resonances can be viewed as standing wave modes trapped within the particle, where the internal electromagnetic field is significantly enhanced. The specific values of \(x\) and \(n\) required to excite these resonances (corresponding to peaks in the Mie coefficients \(|a_l|^2\) or \(|b_l|^2\)) are highly sensitive to \(n\) \cite{ref7,ref11}.

As the real part \(n\) increases, the optical size of the particle effectively increases relative to the surrounding medium. This causes the particle's resonance spectrum to shift, bringing different multipole modes into or out of resonance for a fixed size parameter \(x\). The complex, oscillatory behavior of Q$_{\text{sca}}$ as \(n\) varies (seen in typical Mie plots in the resonance regime) is a direct manifestation of these successive resonances being excited. Generally, for non-absorbing or weakly absorbing particles with sufficient contrast, increasing the real refractive index leads to stronger coupling to the incident field and enhanced excitation of these resonant modes, resulting in higher total scattered energy, hence increased Q$_{\text{sca}}$ \cite{ref1,ref7}.

Simultaneously, the \emph{nature} of the excited resonances and the resulting interference patterns between the fields scattered by different multipoles determine the shape of the phase function. For size parameters in the resonance regime and with sufficient contrast, scattering is dominated by interference effects. Increasing \(n\) often leads to the excitation or strengthening of lower-order resonances (like electric and magnetic dipole/quadrupole modes) which tend to produce scattering patterns with prominent forward lobes due to constructive interference in that direction \cite{ref11,ref16}. While higher-order modes also exist, the overall effect of increasing \(n\) in this regime often shifts the \emph{balance} of contributions towards more directional scattering patterns. This increased directionality, characterized by a phase function that deviates strongly from isotropy, corresponds to a decrease in the complexity parameter $\widetilde{C}_{\text{p}}$ (or an increase in \(1-\widetilde{C}_{\text{p}}\)).

The synchronization between increasing Q$_{\text{sca}}$ and increasing (1-$\widetilde{C}_{\text{p}}$) when \(n\) increases arises because the conditions that favor stronger overall scattering (excitation of resonances) also tend to favor a shift towards more directional phase functions (driven by the specific angular patterns of the excited modes and their interference). \textbf{For single spheres, the detailed way in which different multipole contributions \(a_l, b_l\) evolve with \(n\) and interfere to form the phase function, while simultaneously determining Q$_{\text{sca}}$, results in such a tightly coupled, almost identical pattern of change that Q$_{\text{sca}}$ and (1-$\widetilde{C}_{\text{p}}$) appear to overlap.} This near-perfect synchronization suggests that in this regime, Q$_{\text{sca}}$ and (1-$\widetilde{C}_{\text{p}}$) are not independent parameters but rather reflect, in slightly different mathematical forms, the same underlying variations in the collective behavior of the excited resonant modes. The alternative $\widetilde{C}_{\text{p}}$ parameter, by equally considering the absolute magnitude of each Legendre component's contribution to the phase function, effectively captures this transition from a less directional to a more directional scattering pattern as the resonance landscape changes with \(n\). This coupling is less dominant when the refractive index contrast is low, as resonance effects are suppressed \cite{ref1,ref7}.

\subsection{Imaginary Part Effect: Resonance Damping and Complexity Reduction}
\label{subsec:imaginary_part_mechanism}

The imaginary part of the refractive index (\(k\)) quantifies the absorption of light within the particle. Absorption directly removes energy from the electromagnetic field, leading to a decrease in the total scattered energy, thus reducing Q$_{\text{sca}}$ \cite{ref1,ref7}. This effect is present regardless of contrast, but its impact relative to scattering is important.

Absorption also profoundly impacts the particle's resonance spectrum and the resulting phase function. Electromagnetic resonances within the particle involve light making multiple passes through the material via internal reflections. When the material is absorbing (large \(k\)), the field amplitude decays exponentially during propagation inside the particle according to Beer's law \cite{ref1}. This damping effect is cumulative over the path length. Resonant modes that involve longer internal path lengths, or require multiple internal reflections to build up (often associated with higher-order modes), are much more susceptible to absorption than modes that involve shorter paths or are dominated by surface interactions \cite{ref15}. \textbf{Sufficient refractive index contrast is still important initially to allow for the formation of these internal fields that can then be damped effectively by absorption.}

Higher-order resonant modes and complex interference patterns contributing to the fine angular structure and isotropy of the phase function often rely on such longer internal paths and multiple scattering events within the particle \cite{ref16,ref22}. As \(k\) increases, these higher-order, complex contributions are selectively and strongly damped. This preferential suppression simplifies the resulting scattering phase function, reducing its fine structure and isotropy. A less complex, less isotropic phase function corresponds to a lower complexity parameter $\widetilde{C}_{\text{p}}$.

The synchronization between decreasing Q$_{\text{sca}}$ and decreasing $\widetilde{C}_{\text{p}}$ when \(k\) increases is thus explained by the dual effect of absorption: it reduces the total scattered energy (lowering Q$_{\text{sca}}$) while simultaneously ``eroding'' the complex features of the phase function by selectively damping the resonant modes and interaction pathways responsible for that complexity (lowering $\widetilde{C}_{\text{p}}$). The alternative $\widetilde{C}_{\text{p}}$ parameter is sensitive to this overall simplification of the phase function shape as the resonance landscape is damped by increasing \(k\). This coupling is strongest when absorption becomes a dominant factor in energy loss within the particle's internal field distribution, which is most relevant when these fields are initially significant (i.e., with sufficient refractive index contrast).

\subsection{A Unified Perspective: Resonance Control by n vs.~k, Dependent on Contrast}
\label{subsec:unified_perspective}

In summary, the dual synchronization phenomenon reflects the fundamental distinction between how the real and imaginary parts of the refractive index manipulate the particle's electromagnetic resonance spectrum and the resulting multipole interference patterns. This distinct behavior is most prominent when the \textbf{refractive index contrast is high enough to support well-defined internal fields and resonance phenomena}.

\begin{itemize}
    \item Increasing the real part (\(n\)) \textbf{perturbs} the resonance positions and strengthens their excitation through enhanced phase contrast, leading to higher Q$_{\text{sca}}$ and a shift towards more directional scattering patterns (lower $\widetilde{C}_{\text{p}}$ / higher 1-$\widetilde{C}_{\text{p}}$). The synchronization Q$_{\text{sca}}$ $\propto$ (1-$\widetilde{C}_{\text{p}}$) captures how the \textbf{broad excitation and tuning} of resonant modes by increasing \(n\) is linked to the \textbf{directional characteristics} these modes collectively impart to the phase function.
    \item Increasing the imaginary part (\(k\)) \textbf{dampens} the resonances, particularly those involving longer internal paths, leading to lower Q$_{\text{sca}}$ and a simplification of the phase function (lower $\widetilde{C}_{\text{p}}$). The synchronization Q$_{\text{sca}}$ $\propto$ $\widetilde{C}_{\text{p}}$ captures how the \textbf{loss of energy} due to absorption (reducing Q$_{\text{sca}}$) is coupled to the \textbf{loss of complexity} in the scattering pattern (reducing $\widetilde{C}_{\text{p}}$) as the more complex resonant pathways are suppressed.
\end{itemize}

Our alternative $\widetilde{C}_{\text{p}}$ parameter effectively acts as a metric that is sensitive to how the overall angular distribution, shaped by the superposition of resonant multipole fields, is modified by changes in \(n\) and \(k\). Its formulation, which equally weights the absolute contributions of Legendre polynomials, appears particularly adept at capturing these fundamental shifts in the resonance-driven scattering behavior. This universality across configurations \emph{in the resonance regime} underscores that these mechanisms operate at the level of light-matter interaction governed by the material's optical constants and the particle's size relative to the wavelength, fundamentally affecting the particle's resonant response, provided the interaction strength (contrast) is sufficient. The disappearance of this strong synchronization at low contrast further supports its link to the more complex, resonance-dominated scattering physics rather than simple Rayleigh-like scattering.

% \begin{figure}[htbp]
% \centering
% \includegraphics[width=\linewidth]{figure_mechanism.pdf} % Ensure you have figure_mechanism.pdf in the figures directory
% \caption{Physical mechanisms underlying the dual synchronization behavior. The top panel illustrates how increasing the real part of the refractive index (\(n\)) enhances phase contrast and perturbs resonances, leading to stronger reflection/refraction, increased overall scattering efficiency (Q$_{\text{sca}}$), and a more directional (less isotropic) scattering pattern (lower $\widetilde{C}_{\text{p}}$, higher 1-$\widetilde{C}_{\text{p}}$). The bottom panel shows how increasing the imaginary part (\(k\)) enhances absorption within the particle and dampens resonances, reducing the total scattered energy (Q$_{\text{sca}}$) and preferentially suppressing higher-order scattering paths, resulting in a simplified angular distribution (lower $\widetilde{C}_{\text{p}}$). These represent fundamentally different interactions, most prominent when refractive index contrast is sufficient for resonance effects.}
% \label{fig:mechanisms}
% \end{figure}

\subsection{Mathematical Demonstrability}
\label{subsec:mathematical_demonstrability}

While our numerical results show striking correlation, particularly the near-perfect overlap for single spheres in the real part variation, a rigorous analytical proof of this dual synchronization behavior for general \(m\) and \(x\) in the resonance regime is \textbf{mathematically challenging, likely intractable}. The complex dependence of Mie coefficients on \(m\), the non-linear nature of calculating $\widetilde{C}_{\text{p}}$ (involving integrals and absolute values of sums of complex functions), and the intricate interplay of multiple resonant modes make a closed-form, universally valid proof extremely difficult. Our work relies on high-precision numerical simulation as a powerful tool to discover and verify this complex phenomenon, which is a standard approach in modern scattering research when analytical solutions are not feasible \cite{ref1,ref2,ref3}.

\section{Application Prospects}
\label{sec:application_prospects}

The discovery of the dual synchronization behavior has significant implications for both fundamental research and practical applications in diverse fields leveraging light scattering, particularly within the optical regimes where this phenomenon is prominent.

\subsection{Advanced Optical Material Characterization and Sensing}
\label{subsec:characterization_sensing}

The ability to infer information about both the total scattering strength (Q$_{\text{sca}}$) and the angular distribution complexity ($\widetilde{C}_{\text{p}}$) from a single set of angular scattering measurements is highly valuable. Furthermore, the distinctive synchronization patterns for real versus imaginary refractive index variations offer a novel pathway for material characterization and sensing. This provides a \textbf{diagnostic tool} capable of distinguishing the dominant physical mechanism driving changes in optical properties -- whether it's related to alterations in phase speed (real part, affecting resonance tuning) or absorption (imaginary part, affecting resonance damping) -- by analyzing the specific Q$_{\text{sca}}$-$\widetilde{C}_{\text{p}}$ correlation pattern. This could be particularly useful in applications like distinguishing between particle growth (affecting \(n\) and \(x\)) and surface coating with an absorbing layer (affecting \(k\)) \cite{ref31,ref33}, provided the system remains within the regime where dual synchronization is observed.

\begin{figure}[H]
    \centering
    \includegraphics[width=\linewidth]{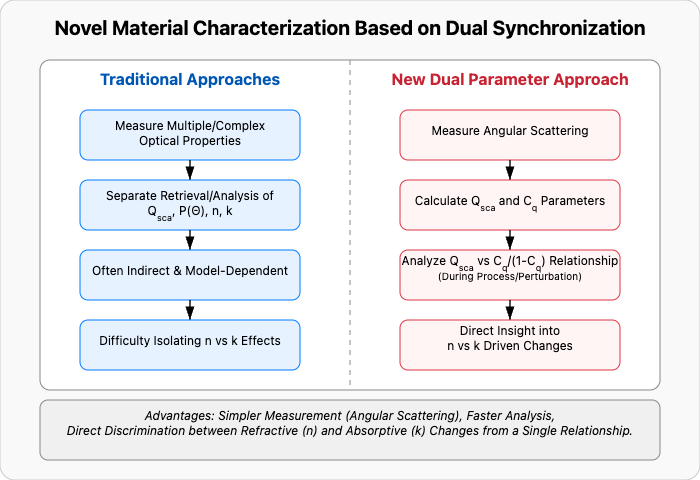} % Ensure you have figure_char_design.pdf in the figures directory
    \caption{Comparison between traditional optical characterization approaches and a potential new method based on monitoring the dual synchronization behavior of Q$_{\text{sca}}$ and $\widetilde{C}_{\text{p}}$. Traditional methods often require complex setups or multiple measurements to retrieve different parameters. The dual synchronization offers the possibility of distinguishing the dominant type of refractive index change (real vs.~imaginary) by observing how Q$_{\text{sca}}$ correlates with $\widetilde{C}_{\text{p}}$ (or 1-$\widetilde{C}_{\text{p}}$), potentially simplifying in-situ or real-time characterization by leveraging the resonance-driven coupling revealed by these parameters \textbf{in the appropriate regime}.}
    \label{fig:char_design}
    \end{figure}

Figure \ref{fig:char_design} illustrates a potential workflow for this new method. This could enable a simplified approach to material characterization and sensing, particularly beneficial in fields dealing with resonant particles (e.g., plasmonics, colloidal optics \cite{ref6,ref27}) or dynamic processes where optical properties change significantly. By measuring the angular scattering intensity of a sample undergoing a change (e.g., swelling, chemical reaction \cite{ref32}, nanoparticle growth or dissolution kinetics \cite{ref33}) at a single wavelength, one can simultaneously calculate Q$_{\text{sca}}$ and the alternative $\widetilde{C}_{\text{p}}$. Observing whether Q$_{\text{sca}}$ tracks (1-$\widetilde{C}_{\text{p}}$) or $\widetilde{C}_{\text{p}}$ directly provides information about whether the dominant change in the material's optical properties is related to altered phase speed (structure/composition change affecting \(n\) and resonance tuning) or altered absorption (introduction/removal of absorbing species affecting \(k\) and resonance damping) \cite{ref15,ref23}, provided the system is in the relevant scattering regime with sufficient contrast.

Examples include assessing the hydration state of biological tissues \cite{ref4}, or detecting the presence of chromophores in solutions. A simple goniometer measuring scattering intensity at various angles at a single wavelength could potentially provide insights previously requiring spectroscopic ellipsometry or complex inverse scattering techniques \cite{ref23}, by specifically exploiting the Q$_{\text{sca}}$-$\widetilde{C}_{\text{p}}$ dual synchronization as a diagnostic indicator in the relevant scattering regime. This approach could complement or simplify existing methods like dynamic light scattering or static light scattering which often focus on size and structure \cite{ref27}.

\subsection{Rational Design of Optical Materials}
\label{subsec:material_design}

The clear understanding of how \(n\) and \(k\) independently steer the coupled behavior of Q$_{\text{sca}}$ and $\widetilde{C}_{\text{p}}$ offers a new design principle for optical materials, particularly those whose function relies on resonance effects. By controlling the complex refractive index, materials can be engineered to achieve specific total scattering efficiencies along with tailored angular distributions, essentially by tuning or damping the underlying electromagnetic resonances and leveraging the coupled Q$_{\text{sca}}$-$\widetilde{C}_{\text{p}}$ response \cite{ref10,ref24}.

\begin{figure}[htbp]
\centering
\includegraphics[width=0.8\linewidth]{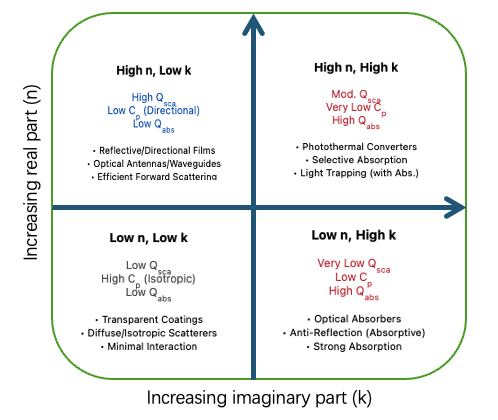} % Ensure you have figure_design_space.pdf in the figures directory
\caption{The material design space visualized through the lens of dual synchronization \textbf{in the resonance regime with sufficient contrast}. Different combinations of high/low real (\(n\)) and imaginary (\(k\)) parts of the refractive index result in distinct regimes of scattering efficiency (Q$_{\text{sca}}$) and complexity parameter ($\widetilde{C}_{\text{p}}$), offering pathways for designing materials with tailored optical functionalities such as directional scattering (High \(n\), Low \(k\)), efficient absorption (High \(k\)), or transparency (Low \(n\), Low \(k\)). These properties arise from the resonance characteristics determined by \(n\) and \(k\). Arrows indicate the conceptual direction of change in properties when increasing \(n\) or \(k\).}
\label{fig:design_space}
\end{figure}

Figure \ref{fig:design_space} illustrates the material design space enabled by this dual synchronization. For example, achieving materials that efficiently scatter light predominantly in a specific direction (e.g., for displays, solar energy harvesting \cite{ref34}, or optical communication) would benefit from materials with high \(n\) and low \(k\). Such materials would reside in the ``High n, Low k'' quadrant of the design space (Figure \ref{fig:design_space}), characterized by high Q$_{\text{sca}}$ and low $\widetilde{C}_{\text{p}}$ (high 1-$\widetilde{C}_{\text{p}}$), reflecting strong, directional resonance excitation. Conversely, materials designed for efficient light absorption, such as in photothermal applications or optical filters, might require a higher \(k\). The resulting optical signature would show low Q$_{\text{sca}}$ and low $\widetilde{C}_{\text{p}}$, corresponding to the ``Low n, High k'' or ``High n, High k'' quadrants, reflecting resonance damping and pattern simplification. The dual synchronization provides a direct link between the material property (\(m\)) and the resulting coupled optical response (Q$_{\text{sca}}$-$\widetilde{C}_{\text{p}}$), enabling a more rational design approach compared to trial-and-error or relying on single parameter tuning.

Furthermore, the principle suggests strategies for designing actively tunable optical materials. If a material's refractive index can be dynamically altered (e.g., via temperature, electric field, or chemical stimuli \cite{ref23,ref35}), the Q$_{\text{sca}}$-$\widetilde{C}_{\text{p}}$ relationship will shift along a trajectory dictated by the relative changes in \(n\) and \(k\), effectively tuning or damping the underlying resonance landscape. By monitoring this trajectory, one could potentially track the material's optical state and tailor its response.

\subsection{Fundamental Theoretical Implications}
\label{subsec:theoretical_implications}

Beyond practical applications, our discovery has significant implications for theoretical understanding of light scattering, particularly in the resonance regime. The distinct behaviors observed for real versus imaginary refractive index variations reveal a fundamental duality in how phase-altering (\(n\)) and energy-dissipating (\(k\)) processes couple the total scattered energy and its angular distribution via their influence on particle resonances. This duality, previously not recognized in this systematic form, provides a new perspective on light-matter interactions. \textbf{The striking degree of synchronization, especially the near-perfect overlap for single spheres in the resonance regime, is a powerful observation suggesting that, under these conditions, Q$_{\text{sca}}$ and $\widetilde{C}_{\text{p}}$ are not merely correlated but are profoundly linked manifestations of the same underlying resonance physics.} While not mathematically identical quantities, their synchronized behavior highlights a deeper, shared dependency on how resonant modes are excited, tuned, and damped by the complex refractive index. This strong coupling is a key finding that warrants further theoretical investigation to understand its mathematical roots, although a full analytical proof is likely intractable (Section \ref{subsec:mathematical_demonstrability}).

The alternative $\widetilde{C}_{\text{p}}$ parameter, by providing an \((2l+1)\) weighting of the absolute contributions of Legendre coefficients, appears to capture a different aspect of phase function variability compared to the standard equally weighted version. While the standard \(C_{\text{p}}\) may be more sensitive to the morphological complexity encoded in the angular oscillations \cite{ref9,ref10}, our $\widetilde{C}_{\text{p}}$ seems to reflect changes in the \textbf{overall balance of multipole contributions and the dominant scattering directions}, which are more directly tied to the bulk optical properties (\(n\) and \(k\)) and their effect on resonance excitation and damping \cite{ref16}. This suggests that traditional analyses based solely on the asymmetry parameter or standard $C_{\text{p}}$ may have overlooked important aspects of the scattering physics related to this fundamental \(n\)-\(k\) duality in the resonance regime. The $\widetilde{C}_{\text{p}}$ parameter, therefore, offers a new and potentially more informative metric for characterizing scattering phase functions in contexts where refractive index variations are primary drivers of optical changes. Further research into its mathematical properties and physical interpretation, particularly how $\sum_{l=0}^{\infty} |(2l + 1)\hat{c}_l|$ relates to the collective strength or distribution of excited multipole modes, is well-warranted.

This work provides a new axis for the analysis of scattering data, complementing traditional parameters like the asymmetry parameter. It highlights the value of examining the \emph{relationship} between different scattering observables, rather than analyzing them in isolation. This insight could lead to more unified theoretical frameworks for understanding and predicting complex scattering behaviors, particularly linking macroscopic observables like Q$_{\text{sca}}$ and $\widetilde{C}_{\text{p}}$ to the microscopic dynamics of internal resonances.

Furthermore, the dual synchronization phenomenon may have implications for inverse scattering problems, where optical measurements are used to infer material properties \cite{ref23}. The distinct correlations associated with real versus imaginary refractive index variations provide an additional constraint that could improve the accuracy and robustness of inverse solutions, allowing for better disentanglement of \(n\) and \(k\) effects from scattering data, especially in the resonance regime. Compared to major prior findings in scattering theory (like Mie theory providing exact solutions for spheres \cite{ref7}, T-matrix generalizing to arbitrary shapes \cite{ref18}, or specific resonance conditions like Kerker effects \cite{ref6}), this discovery identifies a \textbf{general principle governing the coupled response of integrated scattering quantities (total energy and angular complexity) to changes in fundamental material properties (n vs.~k) within the resonance scattering regime and under sufficient contrast}. This principle, revealed by the specific behavior of our alternative $\widetilde{C}_{\text{p}}$ parameter and its dependence on refractive index contrast, represents a novel insight into the underlying physics of scattering.

\section{Conclusion}
\label{sec:conclusion}

In this paper, we have reported the discovery of a previously unrecognized dual synchronization behavior between scattering efficiency (Q$_{\text{sca}}$) and an alternative complexity parameter ($\widetilde{C}_{\text{p}}$) in spherical particle systems. Using rigorous numerical simulations based on the Multiple Sphere T-Matrix method, we demonstrated that this relationship exhibits fundamentally different patterns depending on which component of the refractive index is varied. When increasing the real part, Q$_{\text{sca}}$ synchronizes positively with (1-$\widetilde{C}_{\text{p}}$), while increasing the imaginary part causes Q$_{\text{sca}}$ to synchronize positively with $\widetilde{C}_{\text{p}}$ \cite{ref3,ref12}.

This dual behavior is \textbf{particularly strong in the resonance scattering regime }, where it approaches near-perfect overlap for single spheres, and persists with high correlation for random aggregates. This striking synchronization \textbf{diminishes significantly or disappears at low refractive index contrast}, confirming that it is a phenomenon tied to conditions enabling significant light-matter interaction and the excitation of electromagnetic resonances.

This dual behavior reveals the distinct physical mechanisms by which the real and imaginary components of the refractive index influence light scattering, particularly through their impact on the particle's \textbf{electromagnetic resonances}. The real part primarily affects phase contrast and tunes/excites resonances, enhancing both scattering efficiency and directionality simultaneously. In contrast, the imaginary part induces absorption and dampens resonances, selectively suppressing higher-order scattering processes and thereby reducing both scattering efficiency and phase function complexity \cite{ref1,ref6,ref11}.

The universality of this phenomenon across both single spheres and random aggregates \emph{in the resonance regime and under sufficient contrast} points to its fundamental nature in electromagnetic scattering theory, governed by the intrinsic properties of the material rather than specific geometric arrangements. This discovery not only enhances our theoretical understanding of light-matter interactions in this important scattering regime but also opens new pathways for practical applications. The strong correlations observed could enable simplified optical characterization methods capable of distinguishing between real and imaginary refractive index effects from a single measurement, by analyzing the type of Q$_{\text{sca}}$-$\widetilde{C}_{\text{p}}$ synchronization in the relevant regime \cite{ref9,ref17,ref23}.

Future research should focus on experimental validation of these numerical predictions, particularly using well-controlled particle systems with variable complex refractive indices in the resonance regime. Extension of this investigation to non-spherical particles with specific shapes \cite{ref15,ref16,ref36} and strongly coupled systems like dense aggregates or ordered arrays \cite{ref14,ref30} would further test the universality of the dual synchronization phenomenon and explore how shape and near-field interactions modify the resonance landscape and its coupling to Q$_{\text{sca}}$ and $\widetilde{C}_{\text{p}}$ \cite{ref19}. Additionally, \textbf{systematically mapping the Q$_{\text{sca}}$-$\widetilde{C}_{\text{p}}$ relationship across different size parameter regimes and analyzing its behavior at the boundary of the high/low contrast regimes} would provide further insights into the applicability and limitations of this discovery \cite{ref10,ref20}. Exploring potential approximate analytical approaches to describe this synchronization in simplified scenarios could also be a valuable direction, alongside further fundamental study of the mathematical properties and physical meaning of the \(\sum_{l=0}^{\infty} |(2l + 1)\hat{c}_l|\) quantity itself.

The alternative $\widetilde{C}_{\text{p}}$ parameter introduced in this work provides a new dimension for analyzing scattering properties, complementing traditional metrics such as the asymmetry parameter or the standard complexity parameter. Its unique sensitivity to refractive index variations, specifically related to how it captures the overall balance of multipole contributions affected by resonance tuning and damping, suggests potential applications in designing materials with precisely controlled optical responses and in developing more sophisticated light scattering models and inverse problem algorithms \cite{ref3,ref6,ref23}.

In summary, the dual synchronization relationship between scattering efficiency and phase function complexity, observed prominently in the resonance scattering regime with sufficient refractive index contrast, represents a fundamental optical principle with significant theoretical implications and practical applications. This discovery contributes to our evolving understanding of light scattering phenomena and provides new tools for optical materials analysis and design.

% --- Appendix ---
\appendix
\section{Derivation of Relationships for Complexity Parameters}
\label{app:derivations}

This appendix provides the detailed mathematical derivations for the relationships between the complexity parameters (\(C_{\text{p}}\) and \(\widetilde{C}_{\text{p}}\)) and the asymmetry parameter (\(g\)) for the Henyey-Greenstein (HG) phase function.

\subsection{Relationship between \(\widetilde{C}_{\text{p}}\) and \(g\) for the HG Function}

The Henyey-Greenstein phase function is defined by Legendre expansion coefficients \(\hat{c}_l = g^l\), where \(g\) is the asymmetry parameter and \(|g| < 1\). The alternative complexity parameter \(\widetilde{C}_{\text{p}}\) is defined as:
\begin{align*}
    \widetilde{C}_{\text{p}} = \left( \sum_{l=0}^{\infty} |(2l + 1)\hat{c}_l| \right)^{-1}
\end{align*}
Substituting \(\hat{c}_l = g^l\) and taking the absolute value gives:
\begin{align*}
    \widetilde{C}_{\text{p}} = \left( \sum_{l=0}^{\infty} (2l + 1)|g|^l \right)^{-1}
\end{align*}
Let \(S = \sum_{l=0}^{\infty} (2l + 1)x^l\), where \(x = |g|\) and \(0 \le x < 1\). We can write out the series:
\begin{align*}
    S = 1 + 3x + 5x^2 + 7x^3 + \dots
\end{align*}
Multiplying by \(x\):
\begin{align*}
    xS = x + 3x^2 + 5x^3 + 7x^4 + \dots
\end{align*}
Subtracting the second series from the first:
\begin{align*}
    S - xS &= 1 + (3x-x) + (5x^2 - 3x^2) + (7x^3 - 5x^3) + \dots \\
    (1-x)S &= 1 + 2x + 2x^2 + 2x^3 + \dots \\
    (1-x)S &= 1 + 2x(1 + x + x^2 + \dots)
\end{align*}
The term in the parenthesis is a standard geometric series with sum \(\frac{1}{1-x}\). Therefore:
\begin{align*}
    (1-x)S &= 1 + 2x \left( \frac{1}{1-x} \right) \\
    (1-x)S &= \frac{1-x+2x}{1-x} = \frac{1+x}{1-x}
\end{align*}
Solving for \(S\):
\begin{align*}
    S = \frac{1+x}{(1-x)^2} = \frac{1+|g|}{(1-|g|)^2}
\end{align*}
Since \(\widetilde{C}_{\text{p}} = S^{-1}\), we arrive at the final relationship:
\begin{align}
    \widetilde{C}_{\text{p}} = \frac{(1-|g|)^2}{1+|g|}
\end{align}
From this, we can derive the expression for \(1-\widetilde{C}_{\text{p}}\):
\begin{align*}
    1 - \widetilde{C}_{\text{p}} &= 1 - \frac{(1-|g|)^2}{1+|g|} \\
    &= \frac{1+|g|}{1+|g|} - \frac{1 - 2|g| + |g|^2}{1+|g|} \\
    &= \frac{(1+|g|) - (1 - 2|g| + |g|^2)}{1+|g|} \\
    &= \frac{1 + |g| - 1 + 2|g| - |g|^2}{1+|g|} \\
    &= \frac{3|g| - |g|^2}{1+|g|} = \frac{|g|(3-|g|)}{1+|g|}
\end{align*}

\subsection{Relationship between \(\widetilde{C}_{\text{p}}\) and \(C_{\text{p}}\) for the HG Function}
We start with the two known relationships for the HG phase function:
\begin{enumerate}
    \item \(C_{\text{p}} + |g| = 1\), which implies \(C_{\text{p}} = 1 - |g|\) and \(|g| = 1 - C_{\text{p}}\).
    \item \(\widetilde{C}_{\text{p}} = \frac{(1-|g|)^2}{1+|g|}\), derived above.
\end{enumerate}
Our goal is to express \(\widetilde{C}_{\text{p}}\) in terms of \(C_{\text{p}}\) by eliminating \(|g|\). We can directly substitute the expressions from (1) into (2).

The numerator of the expression for \(\widetilde{C}_{\text{p}}\) is \((1-|g|)^2\). Using \(1-|g| = C_{\text{p}}\), this becomes:
\begin{align*}
    \text{Numerator} = C_{\text{p}}^2
\end{align*}
The denominator is \(1+|g|\). Using \(|g| = 1 - C_{\text{p}}\), this becomes:
\begin{align*}
    \text{Denominator} = 1 + (1 - C_{\text{p}}) = 2 - C_{\text{p}}
\end{align*}
Combining the new numerator and denominator gives the direct relationship:
\begin{align}
    \widetilde{C}_{\text{p}} = \frac{C_{\text{p}}^2}{2 - C_{\text{p}}}
\end{align}
This equation provides a direct mapping between the two complexity parameters for any system that can be described by the Henyey-Greenstein phase function.

\section{Proof of the Legendre Coefficient Bound \(|\hat{c}_l| \le 1\)}
\label{app:coeff_bound}

This appendix provides a proof that for any physically realistic, normalized scattering phase function, the magnitude of its Legendre expansion coefficients is bounded by 1.

A physical scattering phase function \(P(\mu)\), where \(\mu = \cos\Theta\), must satisfy two fundamental conditions:
\begin{enumerate}
    \item \textbf{Non-negativity:} \(P(\mu) \ge 0\) for all \(\mu \in [-1, 1]\), as it represents a probability distribution for scattered intensity.
    \item \textbf{Normalization:} \(\frac{1}{2}\int_{-1}^{1} P(\mu) d\mu = 1\).
\end{enumerate}

The Legendre coefficients \(\hat{c}_l\) are defined by the integral based on the orthogonality of Legendre polynomials:
\begin{align*}
    \hat{c}_l = \frac{1}{2} \int_{-1}^{1} P(\mu) P_l(\mu) d\mu
\end{align*}

To prove the bound, we take the absolute value of the definition:
\begin{align*}
    |\hat{c}_l| = \left| \frac{1}{2} \int_{-1}^{1} P(\mu) P_l(\mu) d\mu \right|
\end{align*}

Using the triangle inequality for integrals, which states that \(|\int f(x) dx| \le \int |f(x)| dx\), we get:
\begin{align*}
    |\hat{c}_l| \le \frac{1}{2} \int_{-1}^{1} |P(\mu) P_l(\mu)| d\mu = \frac{1}{2} \int_{-1}^{1} |P(\mu)| |P_l(\mu)| d\mu
\end{align*}

From the non-negativity condition of the phase function, we know that \(|P(\mu)| = P(\mu)\). The inequality simplifies to:
\begin{align*}
    |\hat{c}_l| \le \frac{1}{2} \int_{-1}^{1} P(\mu) |P_l(\mu)| d\mu
\end{align*}

A key property of Legendre polynomials \(P_l(\mu)\) is that their absolute value is bounded by 1 on the interval \(\mu \in [-1, 1]\):
\begin{align*}
    |P_l(\mu)| \le 1 \quad \text{for all } \mu \in [-1, 1]
\end{align*}

By substituting this upper bound for \(|P_l(\mu)|\) into our inequality, we obtain:
\begin{align*}
    |\hat{c}_l| \le \frac{1}{2} \int_{-1}^{1} P(\mu) \cdot (1) d\mu = \frac{1}{2} \int_{-1}^{1} P(\mu) d\mu
\end{align*}

The final integral, \(\frac{1}{2} \int_{-1}^{1} P(\mu) d\mu\), is precisely the normalization condition for the phase function, which equals 1. Therefore, we arrive at the desired result:
\begin{align}
    |\hat{c}_l| \le 1
\end{align}
This fundamental bound holds for all expansion orders \(l \ge 0\).

\section{Proof of Monotonicity for the Evolving Complexity Parameter \(C_{\text{p}}(s)\)}
\label{app:monotonicity_proof}

This appendix provides the proof that the evolving Symmetry Parameter \(C_{\text{p}}(s) = \left(\sum_{l=0}^{\infty} |\hat{c}_l|^s\right)^{-1}\) is a monotonically increasing function of the scattering order \(s\) for any anisotropic phase function, as stated in Section \ref{subsec:complexity_parameter}.

Let us define the normalization factor \(Z_{\text{scat}}(s) = \sum_{l=0}^{\infty} |\hat{c}_l|^s\), such that \(C_{\text{p}}(s) = 1/Z_{\text{scat}}(s)\). Differentiating \(Z_{\text{scat}}(s)\) with respect to \(s\) yields:
\begin{align*}
    \frac{dZ_{\text{scat}}(s)}{ds} = \sum_{l=0}^{\infty} |\hat{c}_l|^s \log|\hat{c}_l|
\end{align*}
We then define the Mean Logarithmic Stability \(M(s)\) as the expectation value of \(\log|\hat{c}_l|\):
\begin{align*}
    M(s) = \frac{\sum_{l=0}^{\infty} |\hat{c}_l|^s \log|\hat{c}_l|}{Z_{\text{scat}}(s)}
\end{align*}
This allows us to write \(\frac{dZ_{\text{scat}}(s)}{ds} = Z_{\text{scat}}(s) M(s)\). Now, we differentiate \(C_{\text{p}}(s)\) with respect to \(s\):
\begin{align*}
    \frac{dC_{\text{p}}(s)}{ds} & = \frac{d}{ds}\left(\frac{1}{Z_{\text{scat}}(s)}\right) \\
   &  = -\frac{1}{[Z_{\text{scat}}(s)]^2}\frac{dZ_{\text{scat}}(s)}{ds} \\
    & = -[C_{\text{p}}(s)]^2 [Z_{\text{scat}}(s) M(s)] \\
    & = -C_{\text{p}}(s) M(s)
\end{align*}

To determine the sign of the derivative, we analyze the signs of \(C_{\text{p}}(s)\) and \(M(s)\):
\begin{itemize}
    \item \textbf{Sign of \(C_{\text{p}}(s)\):} As \(Z_{\text{scat}}(s) = \sum_{l=0}^{\infty} |\hat{c}_l|^s \geq |\hat{c}_0|^s = 1\), \(C_{\text{p}}(s)\) is always positive for any \(s\).
    \item \textbf{Sign of \(M(s)\):} For any physical phase function, \(|\hat{c}_l| \le 1\), which implies \(\log|\hat{c}_l| \le 0\). The \(l=0\) term in the sum for \(M(s)\) is zero because \(\log|\hat{c}_0|=\log(1)=0\). For any anisotropic phase function, there is at least one \(l' \ge 1\) such that \(|\hat{c}_{l'}| < 1\), which makes \(\log|\hat{c}_{l'}| < 0\). Since all terms in the sum for \(M(s)\) are non-positive and at least one term is strictly negative (for \(s>0\)), \(M(s)\) must be strictly negative for any anisotropic scattering.
\end{itemize}
Combining the signs, we find that for any anisotropic scattering:
\begin{align*}
    \frac{dC_{\text{p}}(s)}{ds} = \underbrace{-C_{\text{p}}(s)}_{<0} \underbrace{M(s)}_{<0} > 0
\end{align*}
For the case of purely isotropic scattering, \(M(s)=0\), which results in \(\frac{dC_{\text{p}}(s)}{ds} = 0\). Therefore, for any physical scattering process, the relationship \(\frac{dC_{\text{p}}(s)}{ds} \geq 0\) holds. This rigorously proves that \(C_{\text{p}}(s)\) is a monotonically non-decreasing function of the scattering order \(s\).

\end{document}